\documentclass[fleqn,usenatbib]{mnras}

\usepackage[T1]{fontenc}

\DeclareRobustCommand{\VAN}[3]{#2}
\let\VANthebibliography\thebibliography
\def\thebibliography{\DeclareRobustCommand{\VAN}[3]{##3}\VANthebibliography}

\usepackage[usenames,dvipsnames]{color}
\usepackage{graphicx}	
\usepackage{amsmath}	
\usepackage{amssymb}	
\usepackage{enumerate}
\usepackage{enumitem}
\usepackage{siunitx}
\usepackage{CJK}
\usepackage{url}
\usepackage{anyfontsize}

\newcommand{\Sref}[1]{Section \ref{#1}}
\newcommand{\Tref}[1]{Table \ref{#1}}
\newcommand{\Aref}[1]{Appendix \ref{#1}}
\sfcode`\.=1001\sfcode`\?=1001\sfcode`\!=1001
\newcommand{\Fref}[1]{\ifhmode \ifnum\spacefactor=1001 Figure \ref{#1}\else Fig.\ \ref{#1}\fi \else Figure \ref{#1}\fi}
\newcommand{\Eref}[1]{\ifhmode \ifnum\spacefactor=1001 Equation (\ref{#1})\else equation (\ref{#1})\fi \else Equation (\ref{#1})\fi}

\DeclareSIUnit\angstrom{\mbox{\normalfont\AA}}

\defcitealias{Kordopatis2015}{K15}

\title[Probing the strength of radial migration]{Probing the strength of radial migration via churning by using metal-rich red giant stars from APOGEE}

\author[C. Lehmann et al.]{Christian Lehmann,$^{1}$\thanks{E-mail: christian.lehmann@geol.lu.se}, Sofia Feltzing$^{1}$, Diane Feuillet$^{1}$, Georges Kordopatis$^{2}$
\\
$^{1}$Lund Observatory, Department of Geology, S\"{o}lvegatan 12, SE-22362 Lund, Sweden \\
$^{2}$Universit\'e C\^ote d'Azur, Observatoire de la C\^ote d'Azur, CNRS, Laboratoire Lagrange, 06000 Nice, France
}

\date{Accepted XXX. Received YYY; in original form ZZZ}

\pubyear{2024}

\begin{document}
\begin{CJK*}{UTF8}{gbsn}
\label{firstpage}
\pagerange{\pageref{firstpage}--\pageref{lastpage}}
\maketitle

\begin{abstract}
Making use of the APOGEE\,DR17 catalogue with high quality data for 143,509 red giant branch stars we explore the strength of different mechanisms that causes a star to radially migrate in the Milky Way stellar disk. At any position in the disk we find stars that are more metal-rich than the local interstellar medium. This is surprising and normally attributed to the migration of these stars after their formation inside their current Galactocentric-radius. Such stars are prime candidates for studying the strength of different migratory processes.
We specifically select two types of metal-rich stars: i) super metal-rich stars ($\mathrm{[Fe/H]}>0.2$) and ii) stars that are more metal-rich than their local environment. For both, we explore the distribution of orbital parameters and ages as evidence of their migration history.
We find that most super metal-rich stars have experienced some amount of churning as they have orbits with $R_g\gtrsim\SI{5}{kpc}$. Furthermore, about half of the super metal-rich stars are on non-circular orbits ($\mathrm{ecc} > 0.15$) and therefore also have experienced blurring. 
The metallicity of young stars in our sample is generally the same as the metallicity of the interstellar medium, suggesting they have not radially migrated yet. Stars with lower metallicity than the local environment have intermediate to old ages. We further find that super metal-rich stars have approximately the same age distribution at all Galactocentric-radii, which suggests that radial migration is a key mechanism responsible for the chemical compositions of stellar populations in the Milky Way.
\end{abstract}




\section{Introduction}
Stars form from gas that has been enriched by previous generations of stars. The naive expectation then would be that the most metal-rich stars within a specific volume would also be the youngest stars. However, we find stars in the solar neighbourhood that are both older and more metal-rich than the Sun. This remarkable observation has intrigued astronomers since first described \citep[see e.g.,][]{Grenon1972}. Understanding why we find these stars helps us to understand how stars move in the Galaxy. For example, it was found that a net outward migration of stars might be a feature of any disc galaxy \citep{Sellwood2002}. Simulations have shown that this is indeed the case \citep[two early studies being][]{Minchev2006, Roskar2008}. In combination with chemical evolution models that predict that galaxies form inside out, and hence with a radial gradient in metallicity, it indicates that the inner Galaxy is a possible origin for these stars. 

Understanding the formation and evolution of galaxies, especially disc galaxies, is a key topic in modern astrophysics \citep{Kubryk2015, BlandHawthorn2016, Mackereth2017, Vogelsberger2020, Buck2020}. A large fraction of the galaxies are disc galaxies \citep{Nair2010}. The Milky Way is one such disc galaxy, which contains structures we can observe, e.g.\ spiral arms. In the Milky Way, we can use stellar populations to better understand the migration processes that lead to the present day stellar distribution in the disc \citep[e.g. ][]{Sellwood2014, Kordopatis2015, Feltzing2020, ViscasillasVazquez2023}. 

Stars form with the chemical and kinematic signatures of their local environment. The interstellar medium (ISM) closer to the Galactic centre is more chemically enriched \citep[e.g. ][]{ArellanoCordova2020, Myers2022, Magrini2023, Silva2023} because the Galaxy forms more stars inside first (enriching the material) while the outer Galaxy is forming the majority of their stars later. This means that stars forming closer to the Galactic centre generally contain more metals than stars forming at the same time at larger radii.
Similarly, stars adopt the orbits of the gas cloud from which they were formed \citep{Naiman2018, Tacconi2020}. Young stars reflect this as their orbits are most often circular and with a Galactocentric radius close to their birth radius in the Galaxy.
Over time, the kinematics of stars can be influenced by massive structures present in the Galactic disc, such as spiral arms, the bar, and giant molecular clouds \citep[see review of][and references within]{Sellwood2022}. Interactions with these structures can perturb the stars' motions, causing the stars to move away from the circular orbit of their formation. Several processes can contribute to stars moving radially in the stellar disc, resulting in a mix of stars at a given Galactocentric radius that have different chemical signatures from their respective formation radii. Collectively this change of radial position in the Galactic plane is referred to as `radial stellar migration' \citep{Sellwood2002}. 

There are two main processes which can cause radial migration, churning and blurring. \textbf{Churning}: changes the angular momentum/guiding radius of a star due to co-rotation resonances with the pattern speed of the spiral arms or the galactic bar \citep{Sellwood2002, Schoenrich2009a}. This cannot be tracked with stellar kinematics as these stars can maintain a circular orbit after migrating inwards or outwards. \textbf{Blurring}: increases the eccentricity of the orbit of a star, while maintaining the guiding radius, due mainly to Lindblad resonances with Galactic structures \citep{Dehnen2000}. The change in eccentricity is a measurably property of the star's kinematics. 
Additionally, {\bf heating} via interactions with giant molecular clouds can influence the kinematics of a star \citep{Spitzer1951, Sellwood1984, Gustafsson2016, Sharma2021, Fujimoto2023}.
This process is random and increases the dispersion of the star's velocity in all directions (e.g.\ increase of radial velocity dispersion from $\sigma_R\sim\SI{0}{\kilo\metre\per\second}$ to $\approx\SI{70}{\kilo\metre\per\second}$, see \citealt{Mackereth2019}). This can affect stellar orbits in a similar way as blurring.

It is important for our understanding of the evolution of the Milky Way to estimate how strong the churning effect is and how much it influences the present-day distributions of age, metallicity, and Galactocentric radius for the stars in the disc.
Because of radial stellar migration, the chemical composition of stars that have had their orbits changed may not reflect the ISM composition of their current location \citep{Grenon1972,Minchev2018}.
The large-scale chemo-dynamical patterns of populations of a single age change over time as stars get shuffled around within the Milky Way \citep[e.g.][]{Wielen1996, Freeman2002, Kordopatis2015, Frankel2018, Lu2022}.
For example, the radial metallicity  gradient of young stars generally reflects the ISM metallicity gradient, while the radial metallicity gradient of older stars is shallower because older stars had more time to experience radial migration (and therefore have, on average, migrated farther). One would, for example, not expect to find both metal-rich and metal-poor stars with similar ages together in the solar neighbourhood without radial migration \citep{Feng2014,Armillotta2018}. However, the ISM trend may vary even at constant $R$, with azimuthal variations of order $\sim\SI{0.1}{dex}$ \citep{Grand2016, Wenger2019}, which we see as the limitations of the ISM precision that can be measured. 

Super metal-rich (SMR) stars have been studied for a long time, with an early definition in \citet{1969ApJ...157.1279S} and subsequent studies from \citet{Grenon1972} and \citet{BlancVaziaga1973} focused on verifying the SMR measurements. Recently, SMR stars have also been used as tracers for radial stellar migration \citep[e.g.][]{Kordopatis2015, 2023A&A...669A..96D, Nepal2024}.  Metal-rich stars are useful tracers of migration as the ISM metallicity exceeds $\mathrm{[Fe/H]}\sim\SI{0.2}{dex}$ only in the inner regions of the Milky Way \citep[e.g.\ inner Galaxy Cepheids in][]{Silva2023}. Stars with $\mathrm{[Fe/H]}>0.2$ observed at a Galactocentric radius $\gtrsim\SI{5}{kpc}$ must have migrated outwards from the inner Galaxy where they formed. 
If blurring were the only process influencing the motion of these stars, the eccentricities of stars with high metal content ($\mathrm{[Fe/H]}>0.2$) found in the solar neighbourhood ($R\approx\SI{8}{kpc}$) would all exceed $\mathrm{ecc}=0.3$. Any metal-rich stars at 8 kpc with $\mathrm{ecc} < 0.3$ must have experience some amount of churning.

It is seen that a significant fraction of SMR stars found near the solar neighbourhood today have orbital properties consistent with having experience some amount of churning (i.e. have $\mathrm{ecc} < 0.3$). \citet{2023A&A...669A..96D} use 171 metal-rich stars from the Gaia-ESO Survey and find that most of their metal-rich sample has $\mathrm{ecc} < 0.3$. \citet{Kordopatis2015} have a larger sample from RAVE and find that $\sim 50 \%$ of the metal-rich stars have $\mathrm{ecc} < 0.15$. However, this sample only probes $\sim \SI{1.5}{kpc}$ around the Sun. 
With a large sample of red giant stars from APOGEE Data Release 17 \citep[DR17][]{Abdurrouf2022} in combination with data from {\it Gaia} DR3 \citep{2023A&A...674A...1G}, we update previous results on migration of SMR stars throughout the Milky Way disc by probing a larger volume and explore a new selection of local metal-richness.

The APOGEE\,DR17 sample combined with {\it Gaia} provides kinematic and chemical information, which we supplement with age information using the supervised machine learning technique XGBoost \citep{Anders2023} for a sub-sample of these stars. 
We explore the efficiency of blurring and churning processes as a function of distance from the Galactic centre and the dependence on the vertical ($z$) coordinates of individual stars. There is a debate in the literature over whether the efficiency of churning is affected by the distance from the mid-plane of the disc \citep{Minchev2012, Minchev2018, Roskar2013, Halle2015, DiMatteo2015}. Understanding these details is key to quantify the effects of churning and blurring on different populations of stars.

This paper is structured in the following way in Sect.\,\ref{sect:data} we describe the selection of the data used in this project. Section\,\ref{sect:def} sets out the definitions we use, e.g. our working definition of metal-rich stars and Sect.\,\ref{sec:analysis} outlines our main results which are further discussed in Sect.\,\ref{sect:disc}. The paper concludes with a summary in Sect.\,\ref{sec:conclusion}.

\section{APOGEE red giant data catalogue}\label{sec:catalogue}
\label{sect:data}
\begin{table}
\caption{Selection criteria used to create our catalogue from APOGEE~DR17 and \textit{Gaia}. Further details can be found in Sect.\,\ref{sec:catalogue}.}
\begin{center}
\vspace{0cm}
\begin{tabular}{lllll}
 \multicolumn{3}{l}{\textit{APOGEE\,DR17}}\\
  Flag  & \multicolumn{2}{l}{Criterium} \\
\hline
  \tt{TEFF} &$<$& 7000 \\
  \tt{LOGG} &$<$& 2.8  \\
  \tt{SNREV} & $>$ & 80 \\
  \tt{FE\_H\_FLAG} &$=$& 0 \\
  \tt{STARFLAG} & $\neq$ & {\tt VERY\_BRIGHT\_NEIGHBOR}, \\
  & & {\tt PERSIST\_HIGH}, {\tt SUSPECT\_BROAD\_LINES}, \\
  & & or {\tt SUSPECT\_ROTATION} \\
  \tt{ASPCAPFLAG} &$\neq$& {\tt STAR\_BAD}, {\tt CHI2\_BAD}, \\
  & & {\tt M\_H\_BAD}, or {\tt CHI2\_WARN} \\
  \tt{EXTRATARG} & $\neq$ & \tt{DUPLICATE} \\
\hline
  \\
  \multicolumn{3}{l}{\textit{Gaia ERD3}}\\
  Property & \multicolumn{2}{l}{Criterium}\\
\hline
  {$\sigma_{\pi}/\pi$} &  $< $& 0.2\\
\hline
 \end{tabular}
\end{center}
    \label{tab:selection}
\end{table}

We would like to investigate the properties of high metallicity stars in a large volume of the Milky Way, so we chose to use red giant branch stars, which are intrinsically luminous. 
The APOGEE survey is an near-infrared ($H$-band), $R \sim 23000$, stellar spectroscopic survey conducted as part of the third and fourth iterations of the Sloan Digital Sky Survey \citep{2017AJ....154...94M, 2019PASP..131e5001W}.
We select red giant stars from APOGEE DR17 \citep{Abdurrouf2022}, using the selection criteria listed in \Tref{tab:selection}. The selection on stellar atmospheric parameters (`{\tt TEFF}' and `{\tt LOGG}') ensures that the sample consists of only red giant branch stars.
To further ensure high precision in the $\mathrm{[Fe/H]}$ measurements of our sample, we impose a cut on `{\tt SNREV}' to remove low signal-to-noise spectra, which results in $T_\mathrm{eff}$ uncertainties $<\SI{80}{\kelvin}$ and $\mathrm{[Fe/H]}$ uncertainties $<\SI{0.02}{dex}$\footnote{These are internal uncertainties. The real uncertainties (accuracy+precision) are likely higher.}. The `{\tt FE\_H\_FLAG}', `{\tt STARFLAG}', and `{\tt ASCAPFLAG}' flags verify that the data reduction and analysis has worked as intended and no spectra have a high likelihood of light contamination from neighbouring spectra. The criteria imposed on `{\tt EXTRATARG}' returns the higher signal-to-noise ratio entry in the catalogue in case of duplicates. Further information on the meaning of the flags in APOGEE DR17 can be found at their dedicated website \href{https://www.sdss4.org/dr17/data_access/}{\url{https://www.sdss4.org/dr17/data_access/}}. We only select stars with $\sigma_{\pi}/\pi$ to 20\% so that our distance measurements are precise. These selections results in a sample of $143{,}509$ red giant stars.

\begin{figure}
    \centering
    \includegraphics[width=0.445\textwidth]{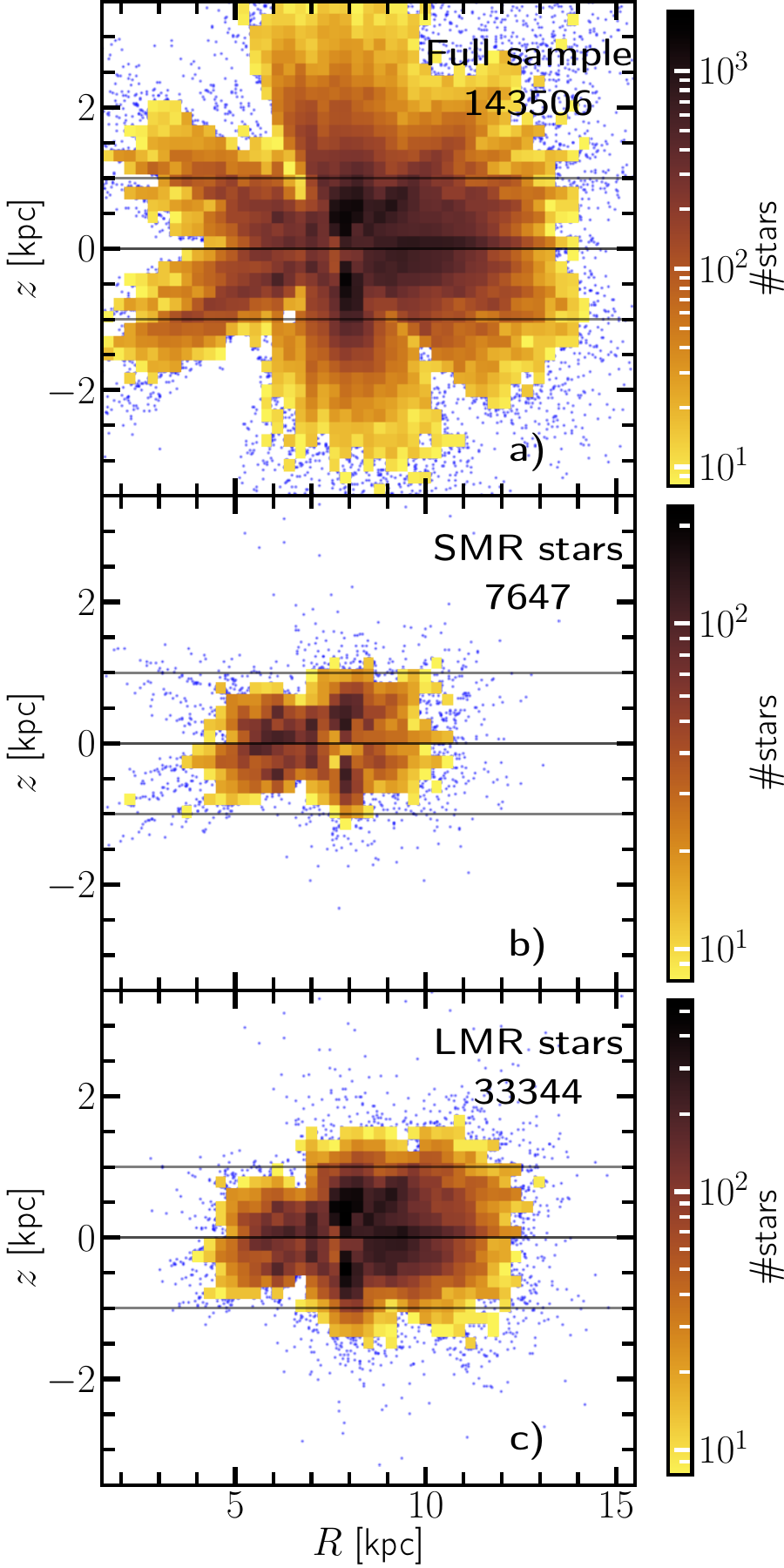}
    \caption{Galactocentric radius vs. distance from the Galactic plane map of our APOGEE DR17 red giant selection. a) the full sample, b) selection of SMR stars, and c) selection of LMR stars. The colour map in each panel represents the number of stars per pixel from blue (few stars) to yellow (many stars).
    The three horizontal lines ($z=-1, 0, 1$) are shown to guide the eye and make easier comparisons between the populations.}
    \label{fig:r_z_map}
\end{figure}

We calculate the positions and kinematics of all stars in the sample following \citet{Kordopatis2023}, using the \textit{Gaia} DR3 RA, dec, and proper motions, geometric distances from \citet[][based on \textit{Gaia} DR3]{BailerJones2021} and APOGEE DR17 spectroscopic radial velocities.  We calculate orbits with \textsc{galpy} \citep{galpy15} and assume the \textsc{actionAngleStaeckel} approximation \citep{Binney2012, Bovy2013} using delta of 0.4 and the potential from \citet{McMillan2017}. Furthermore, we adopt the local standard of rest as $(V_R, V_\phi, V_z)_\odot = (-9.5, 250.7, 8.56) \SI{}{\kilo\metre\per\second}$ \citep{Reid2020} and the solar coordinates as $(R, z)_\odot=(8.249, 0.0208)\SI{}{kpc}$ \citep{GRAVITYCollaboration2020}. The $R-z$ coordinates of our full sample are shown in \Fref{fig:r_z_map}a. For a subset of this sample consisting of $88{,}107$ stars, we also have age estimates available from the work of \citet{Anders2023}. They use XGBoost, a tree-based machine learning algorithm, which achieves uncertainties of $\lesssim10\%$ of individual stellar age. As a machine learning algorithm, the accuracy is dependent on the training set \citep{Miglio2021} which used a combination of spectroscopic and asteroseismic methods to maximise the precision (age spread of $\delta_\mathrm{age}\lesssim\SI{1}{Gyr}$). We inspected this subset of stars for potential biases but other than selecting against stars that are extremely metal-rich ($\mathrm{[Fe/H]}>\SI{0.5}{dex}$) or metal-poor ($\mathrm{[Fe/H]}<\SI{-1}{dex}$), we did not find any.

As our stars are more metal-rich than the Sun, they will have spectra with plethora of absorption lines. Fitting stellar spectra becomes increasingly difficult as the metallicity increases. We have therefore visually inspected approximately 50 spectra of the most metal-rich stars in our sample. We find that the spectra of our stars are well fit by the synthetic model spectra (see  Appendix \ref{app:spectra}).

\section{Definitions}
\label{sect:def}
\subsection{Guiding radius}
\begin{figure}
    \centering
    \includegraphics[width=0.445\textwidth]{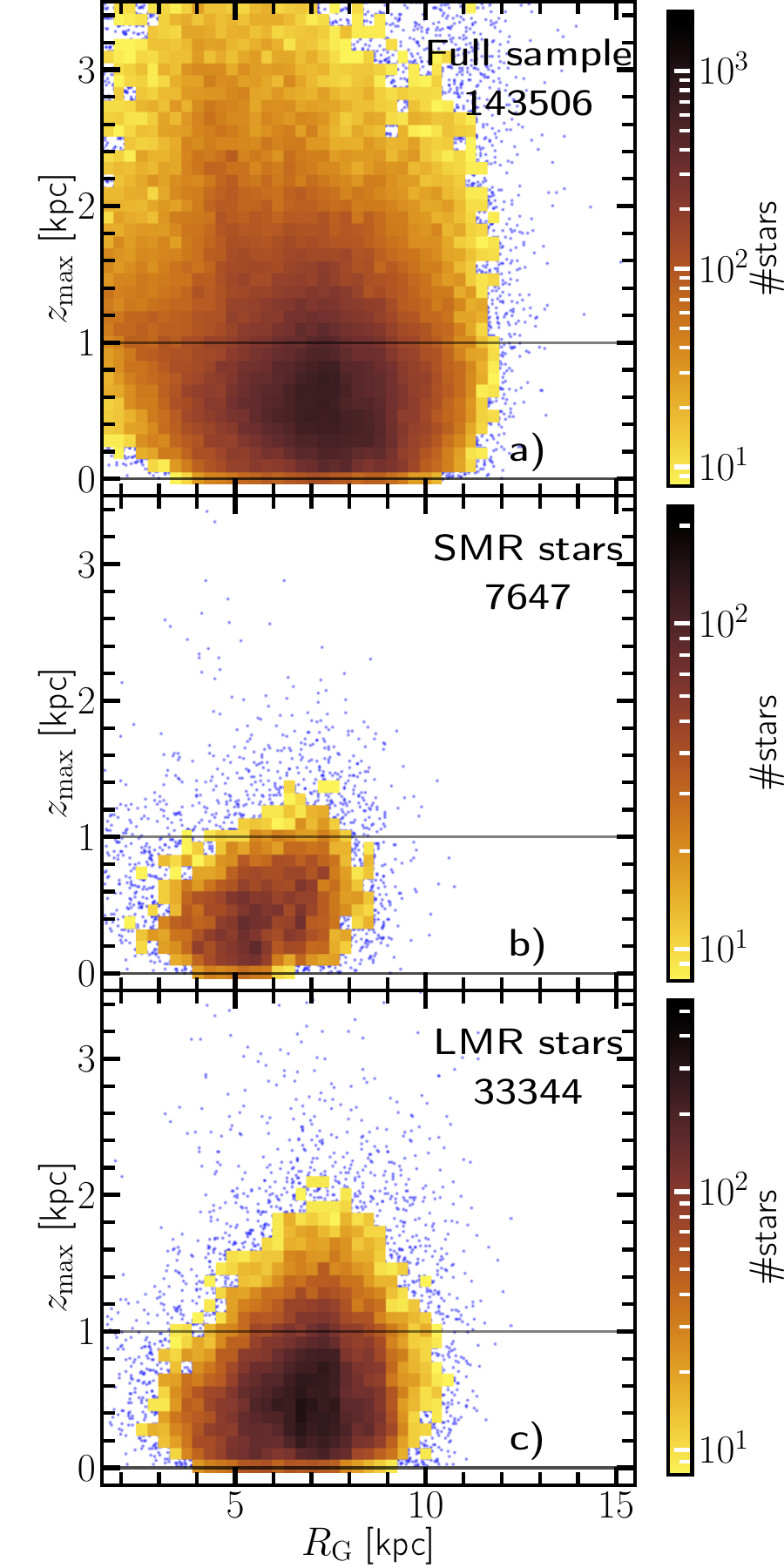}
    \caption{Galactocentric radius vs. distance from the Galactic plane map of our APOGEE DR17 red giant selection. The top panel shows the full sample with the next free panels being different selections in metallicity, as shown in the panel.}
    \label{fig:r_zmax_map}
\end{figure}

The guiding radius $R_\mathrm{G}$ is a characterisation of the orbit of a star. It represents the radius on an imaginary circular orbit that corresponds to the angular momentum $L_\mathrm{z}$ of the star:
\begin{align}
    R_\mathrm{G} = \frac{L_\mathrm{z}}{v_\mathrm{C}(R_\mathrm{G})},
\end{align}
where $v_\mathrm{C}(R_\mathrm{G})$ is the velocity in the direction of the rotation on a circular orbit.
We use the following approximation for the guiding radius in this work:
\begin{align}
    R_\mathrm{G} = \frac{R \cdot v_\phi}{V_\mathrm{C}},
\end{align}
where $R$ and $v_\phi$ are the radius and velocity in the direction of the rotation of the star, and $V_\mathrm{C}$ is the velocity on a circular orbit at $R_\odot=\SI{8.249}{kpc}$ in direction of Galactic rotation (we use $V_\mathrm{C}=V_{\phi, \odot}=\SI{250.7}{\kilo\metre\per\second}$).
We show the guiding radius of our samples  in \Fref{fig:r_zmax_map} together with the maximum height above the plane for the star's orbit ($z_\mathrm{max}$).

\subsection{Metal-rich stars}\label{sec:metal_rich_def}
The presence of stars more metal-rich than the Sun but at similar Galactocentric radii was a surprising finding \citep{Grenon1972}. This caught the attention of astronomers for as long as it has been feasible to measure stellar metallicities. There were worries that the high metallicity measurements would be related to stellar evolutionary effects, which would explain the observations. However, \citet{1999Ap&SS.265..331G} demonstrated this to not be the case. He proposed that the super metal-rich stars we observe in the solar neighbourhood today likely originated in the inner disc of the Galaxy (Fig.\,3 of his paper).

For the purposes of this paper we are interested in using metal-rich stars as tracers of radial migration, which raises the question: how do we define when a star is metal-rich? The purpose of this section is to define metal-richness and discuss if there can be more than one definition that can be used for our purposes. 

There is, to our knowledge, no definitive definition of SMR stars in the literature and often a  metallicity is picked without much of a discussion. Initially, $\mathrm{[Fe/H]} = \SI{0.2}{dex}$ was used to define SMR stars, which is a value based on the measured metallicity of the Hyades stellar cluster \citep{1969ApJ...157.1279S}. But while this is a simple and straightforward definition, the review by \citet{1996ApJS..102..105T} noted that the measured metallicity of the Hyades stellar cluster changes between different analyses and more recent determinations may result in a lower metallicity. \citet{1996ApJS..102..105T} still concludes that $\mathrm{[Fe/H]} = \SI{0.2}{dex}$ is a reasonable value to adopt. He based this recommendation partially on the observation of $\mu$Leo as the most metal-rich giant star known at the time.

Another way to define super metal-richness would be to connect it with the local ISM, i.e.\ a star can be metal-rich compared to its surroundings. This is an appealing possibility that is possible thanks to the large datasets available today. 
We thus propose two different definitions of stars that trace the metal-rich regime of a stellar population:

\begin{enumerate}[wide, labelwidth=!,itemindent=!,labelindent=0pt, leftmargin=0.2cm, itemsep=0.1cm, parsep=0pt]
    \item $\mathrm{[Fe/H]}>$ a certain value for all positions in the Galaxy. We refer to these stars  as Super Metal-Rich (SMR) stars. 
    \item A star is metal-rich if its iron abundance is larger than that of the local ISM at the stars current location. We refer to these stars as Local Metal-Rich  (LMR) stars. 
\end{enumerate}
With the advent of large spectroscopic surveys we are able to obtain much larger datasets of stars and in particular of metal-rich stars.
A relevant question in this context is how we translate between different metallicity scales and between different types of stars, i.e. are stars in RAVE and APOGEE on the same scale, do dwarf stars and giant branch stars with intrinsically the same metallicity have the same derived $\mathrm{[Fe/H]}$? 

\citet{2022A&A...663A...4S} studied the results from six major stellar surveys and a re-analysis of two of them. They compared the results between the surveys and with a "ground truth" catalogue \citep[PASTEL,][]{1980A&AS...41..405C,2010A&A...515A.111S,2016A&A...591A.118S}. 
A common result is that the surveys over-estimated $\mathrm{[Fe/H]}$ for low metallicities while some surveys in addition under-estimated $\mathrm{[Fe/H]}$ at the highest metallicities. All surveys perform well in the well-populated intermediate metallicity range. \citet{2022A&A...663A...4S} Table\,2 summarises the derived $\mathrm{[Fe/H]}$ for three open clusters, M67, Pleiades, and Hyades from five surveys and two re-analyses. The median $\mathrm{[Fe/H]}$ for Hyades varies from $\SI{-0.04}{dex}$ to $\SI{0.2}{dex}$, where APOGEE measured it at $\SI{0.155}{dex}$ (from DR16). The question is then what to chose for our purposes. One option would be to choose the value found for the Hyades in APOGEE or another value that was used in other studies (remembering that the surveys may not necessary have the same scale for their iron abundances). For example, \citet{Kordopatis2015} used $\SI{0.1}{dex}$ and \citet{2023A&A...669A..96D} (\textit{Gaia}-ESO Survey) used $\SI{0.15}{dex}$ as their lower metallicity limits for SMR stars respectively.
\citet{Nepal2024} initially used our second definition \textit{"A SMR star is defined as having a metal abundance that exceeds the metallicity of the local present-day ISM; because of the negative radial abundance gradient in the MW, its value varies with galactocentric distance."} but in the end they appear to use a single value to define SMR stars, $\SI{0.1}{dex}$, and do not give a relation as a function of galactocentric radius.
Therefore, we find little consensus on how to define the exact value to define SMR stars. 
For the purpose of the current study we will adopt the following two definitions: 

\begin{enumerate}[wide, labelwidth=!,itemindent=!,labelindent=0pt, leftmargin=0.2cm, itemsep=0.1cm, parsep=0pt]
    \item $\mathrm{[Fe/H]}> +0.2$ (SMR)
    \item $\mathrm{[Fe/H]} > \mathrm{[Fe/H]}_{\rm ISM}$ (as defined by \Eref{eq:ism}, LMR)
\end{enumerate}

We define the ISM metallicity at any given $R$ with measurements of Cepheids as proxies for the current-day ISM. We can do that, as Cepheids are typically young ($\lesssim\SI{300}{Myr}$, \citealt{Bono2005, Pietrukowicz2021}) and therefore have not moved far from their formation sites and thus are a fair sample of the properties of the gas from which they formed. Specifically, we pick the following relation from \citet{Silva2023} (their Fig. 2 and their Table 4):

\begin{equation}
    \mathrm{[Fe/H]} = (-0.907 \pm 0.046) \log(R) + (0.81 \pm 0.04).
    \label{eq:ism}
\end{equation}

\begin{figure}
    \centering
    \includegraphics[width=0.47\textwidth]{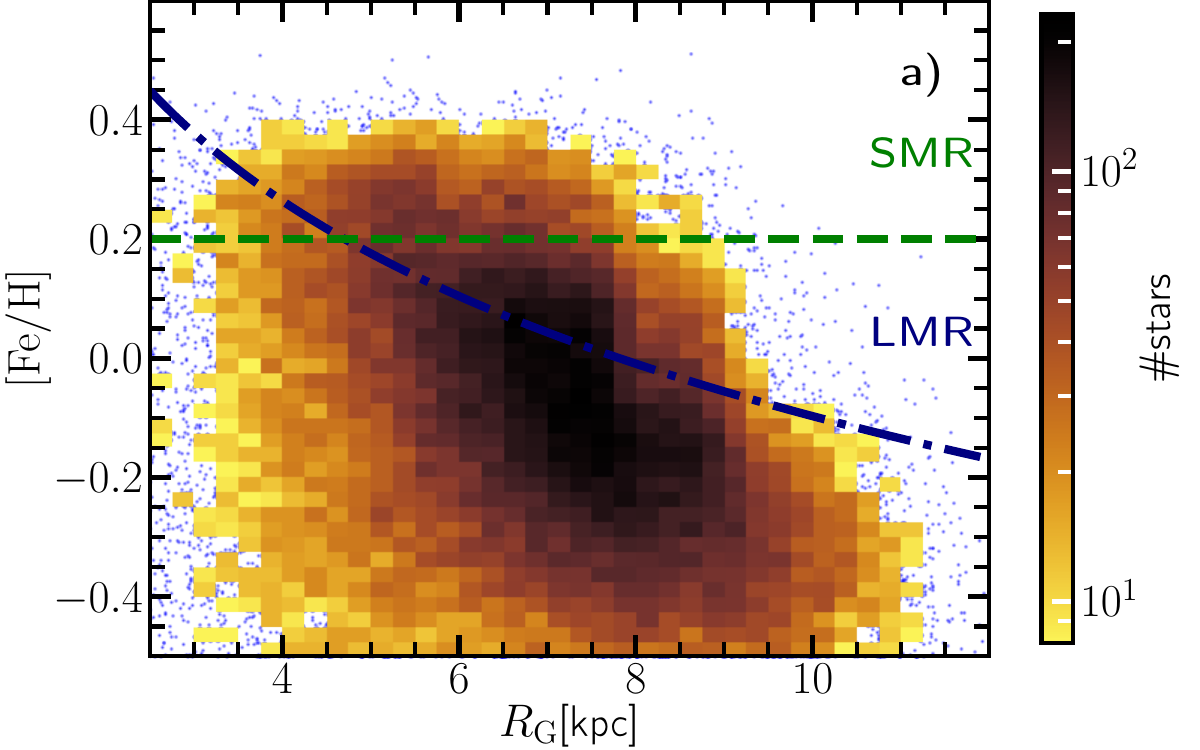}
    \includegraphics[width=0.47\textwidth]{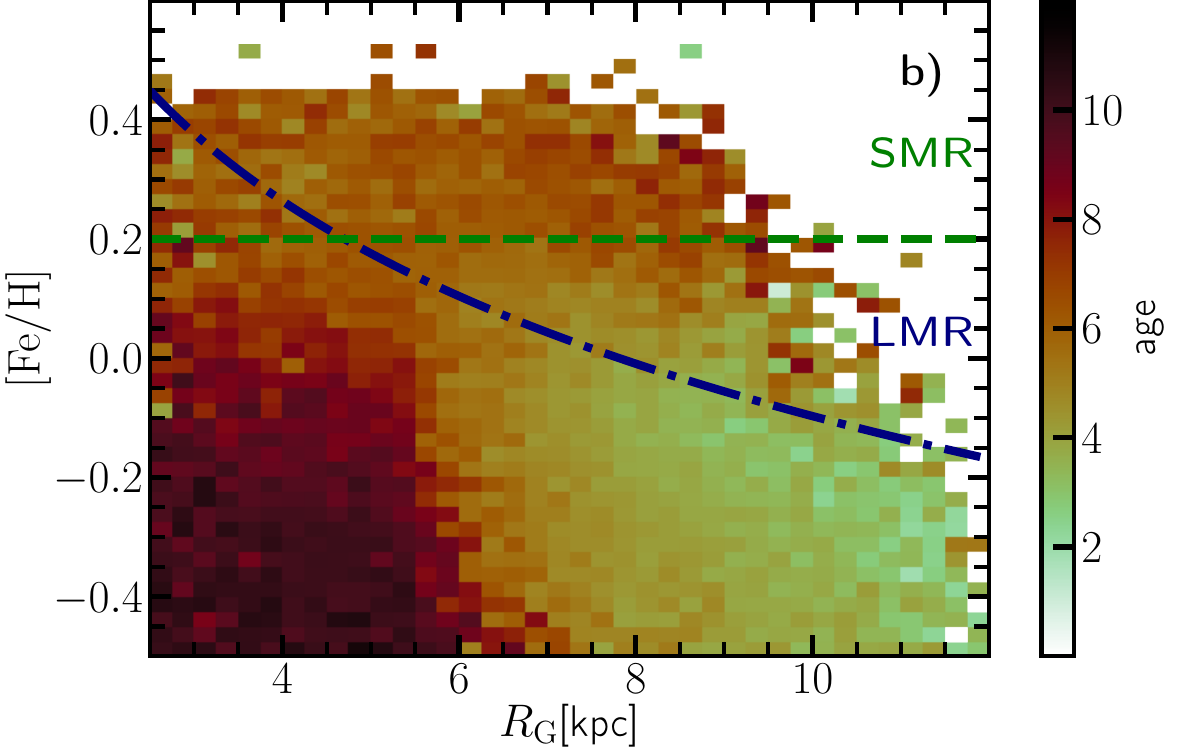}
    \includegraphics[width=0.47\textwidth]{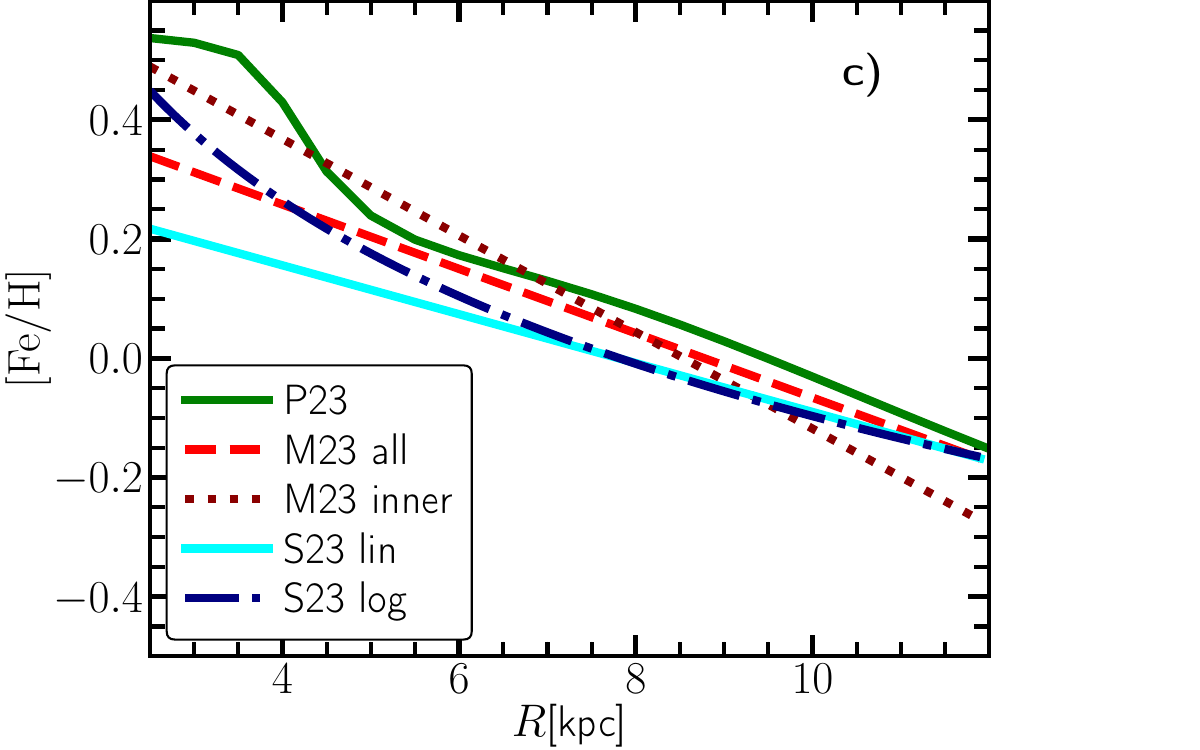}
    \caption{The metallicity-guiding radius diagram for our sample. Stars with $z_\mathrm{max}\leq1$\,kpc colour coded by a) the number of stars in each pixel and b) the median age in each pixel.
    The green dashed line represents the minimum $\mathrm{[Fe/H]}$ for our SMR selection and the dark blue dash dotted line the minimum $\mathrm{[Fe/H]}$ for our LMR selection (see \Sref{sec:metal_rich_def}). c) shows the ISM radial metallicity gradient reported in the literature both measured in the Milky Way and used to model Milky Way chemical evolution. The relations shown were considered for the LMR selection, i.e.\ P23 \citep{Prantzos2023}, M23 \citep{Magrini2023}, and S23 \citep{Silva2023}.}
    \label{fig:R_FeH_age}
\end{figure}

We visualise this selection together with our data set in \Fref{fig:R_FeH_age}. Figure\,\ref{fig:R_FeH_age}c shows the different ISM radial gradients that we considered to use as the definition of the LMR stars. The model by \citet{Prantzos2023} seems to overestimate the metallicity compared to the observational measurements. \citet{Magrini2023} shows curves fitted with open cluster stars (the red dashed line takes into account their whole sample and the dark red dotted line only the inner $\sim \SI{11}{kpc}$) and \citet{Silva2023} shows fits to Cepheid metallicities (the logarithmic fit is shown with the dark blue dash-dotted line and the linear fit is shown in the light blue solid line, \citealt{Silva2023}). We  decided to use the logarithmic Cepheid relationship as simulations suggest non-linearity in the inner Galaxy and since it is in broad agreement with other measurements\footnote{For the innermost regions, the logarithmic measurement is likely an overestimation. But since the majority of our data has $R\gtrsim\SI{4}{kpc}$, we consider it a valid choice.}. Other choices are possible.

The $R-z$ coordinates for sub samples of metal-rich stars (SMR and LMR) are shown in \Fref{fig:r_z_map}\,b and c.
The LMR selection resulted in 19{,}899 stars ($16.5\%$ of the full sample), while the SMR selection has 7{,}647 stars ($6.3\%$ of the full sample) with 7{,}098 stars shared between the selections.

We note that some studies have questioned if very high metallicities are realistic. Notable the series of papers by B. Taylor who on a statistical basis (for the then available data) rule out such high values as +0.3\,dex for $\mu$Leo. Although a relevant critique, we find that it is beyond the scope of the current investigation to delve further in to the potential non-reality of very high metallicities for red giant branch stars. But consider this a valid topic to be investigate further.
Therefore, we did not perform a high-metallicity cut in our sample, as even our most metal-rich star ($\mathrm{[Fe/H]}=0.56$) is compatible (within 4 sigma, considering realistic uncertainties of $\sigma_\mathrm{[Fe/H]}\approx0.04$ from open cluster metallicity variations in \citealt{Donor2018}) with the present-day Bulge metallicity $\mathrm{[Fe/H]}>0.4$ \citep[e.g.][]{ViscasillasVazquez2023, Nepal2024}.

\subsection{Migration}
The process of migration is key to understand the metallicity distribution of stars in the disc. For the purposes of this work, we distinguish between two migration modes, churning and blurring, which we attempt to measure. Additionally, heating can affect radial migration of stars randomly.

\subsubsection{Churning}
Churning is the process that changes the guiding radius (and hence angular momentum) of the star, which in practice will move the star from its original circular orbit to a new circular orbit further inwards/outwards. This migration mode can happen through stars that enter in resonance with the corotation radius of the bar, the spiral arms or overlapping patterns of both \citep[][and references therein]{Sellwood2002, DiMatteo2013, Khoperskov2020, Sellwood2022}.

Churning cannot be tracked via kinematic means alone as stars that migrated via this mechanism have orbits similar to local stars that have not. Therefore, the evidence that points towards migration via churning are stars that are more chemically enriched than their environment, e.g.\ metal-rich stars in the solar neighbourhood.

\subsubsection{Blurring}
Contrary to churning, blurring causes no changes in angular momentum. It changes the movement of stars from relatively circular to eccentric orbits, keeping a constant guiding radius $R_\mathrm{G}$. This can be seen as a migration process as the inner/outer fractions of the orbit now pass through areas of the Milky Way further inwards/outwards the disc that the star would not have entered without blurring \citep{Dehnen1998, Dehnen2000}.
Blurring is mainly caused by stars being in Lindblad resonance with the Galactic bar and therefore experiencing both gravitational pull and push from the bar on different parts of their orbit. Over time, this leads to the orbit being stretched to higher eccentricity.

\subsubsection{Heating}
Observations show that disc stars increase their velocity dispersion the older they are, which was first observed in \citet{Spitzer1951}. This process is called heating and was much explored  \citep[e.g. ][]{Sellwood1984, Gustafsson2016} and increases the velocity dispersion in all directions.
The main source of heating is random interactions of stars with giant molecular clouds \citep{Haenninen2002} and transient spiral structures \citep{DeSimone2004}. Heating does therefore, in effect, seem like a type of blurring (increasing the eccentricity) and also generally increase $z_\mathrm{max}$, which we need to keep in mind when interpreting our results.

\subsubsection{Detecting migration}
Since blurring leaves stars on measurably distinct orbits, this can be traced. However, it appears that churning is necessary to explain the present-day stellar metallicity distribution. For example, a star in the solar neighbourhood ($R=\SI{8}{kpc}$) with $\mathrm{[Fe/H]}=0.1 (0.2)$ would need $\mathrm{ecc}>0.2 (0.3)$ (using the ISM predictions via Cepheids from \citealt{Silva2023}) so that the orbit can be explained by blurring alone. 
As several studies of metal-rich stars in the solar neighbourhood have found \citep[e.g.][]{Kordopatis2015, Sharma2021a}, the majority of metal-rich stars are not on orbits as eccentric as this. Therefore, churning must be a factor in their migration.
So while blurring can be detected (via present-day orbit eccentricities), churning is mostly inferred via the chemistry of stars.

\subsection{Circular orbits -- definition}
We define the stellar orbits as circular if they have $\mathrm{ecc}<0.15$ and as non-circular if they have $\mathrm{ecc}\geq0.15$. Since blurring increases the eccentricity of the orbit and churning preserves the eccentricity of the orbits, we assume stars on circular-orbits have  either stayed at their natal orbit or migrated mainly via churning, while stars on non-circular orbits have experienced blurring. It is, of course, possible that a star is first churned and then blurred. Note, that the exact eccentricity at which to split the orbits into circular and non-circular is arbitrary. We use the same split as \citet[][]{Kordopatis2015}. For context, a star with a pericenter of $R=\SI{8}{kpc}$ and an orbital eccentricity $\mathrm{ecc}=0.15$ has an apocenter of $R=\SI{10.8}{kpc}$, while a stars with an apocenter of $R=\SI{8}{kpc}$ and an orbital eccentricity $\mathrm{ecc}=0.15$ has an pericenter of $R=\SI{5.9}{kpc}$.

\section{Results}\label{sec:analysis}
When focusing on metal-rich stars we are interested in all the processes that potentially affect their migration. Our metal-rich selection is visualised in \Fref{fig:R_FeH_age}~a and b as stars above the SMR and LMR lines, as well as \Fref{fig:r_z_map}~b and c and \Fref{fig:r_zmax_map}, which show the $R$--$z$ and $R_G$--$z_\mathrm{max}$ maps, respectively, of the SMR and LMR selections, respectively. 

\subsection{Circularity of stellar orbits}
\begin{figure}
    \centering
    \includegraphics[width=0.47\textwidth]{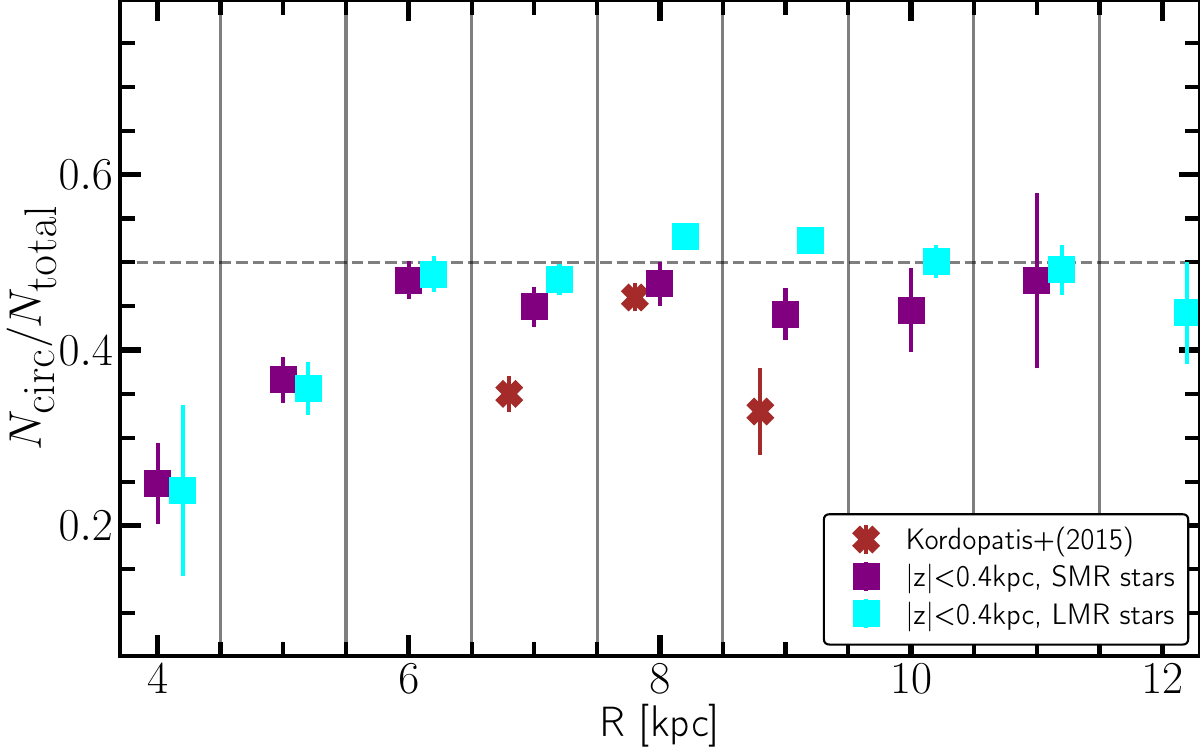}
    \caption{Fraction of stars on circular orbits relative to the total number of stars, split into $R$-bins. Comparison between the values from \citetalias{Kordopatis2015} (brown symbols) and our results for SMR and LMR stars, respectively (purple and cyan symbols). The SMR selection corresponds to the selection done in \citetalias{Kordopatis2015}, i.e. the combination of the [0.2, 0.3]\;dex and [0.3, 0.4]\;dex metallicity bins in their Table 3.}
    \label{fig:Kord_comparison}
\end{figure}

\begin{figure}
    \centering
    \includegraphics[width=0.47\textwidth]{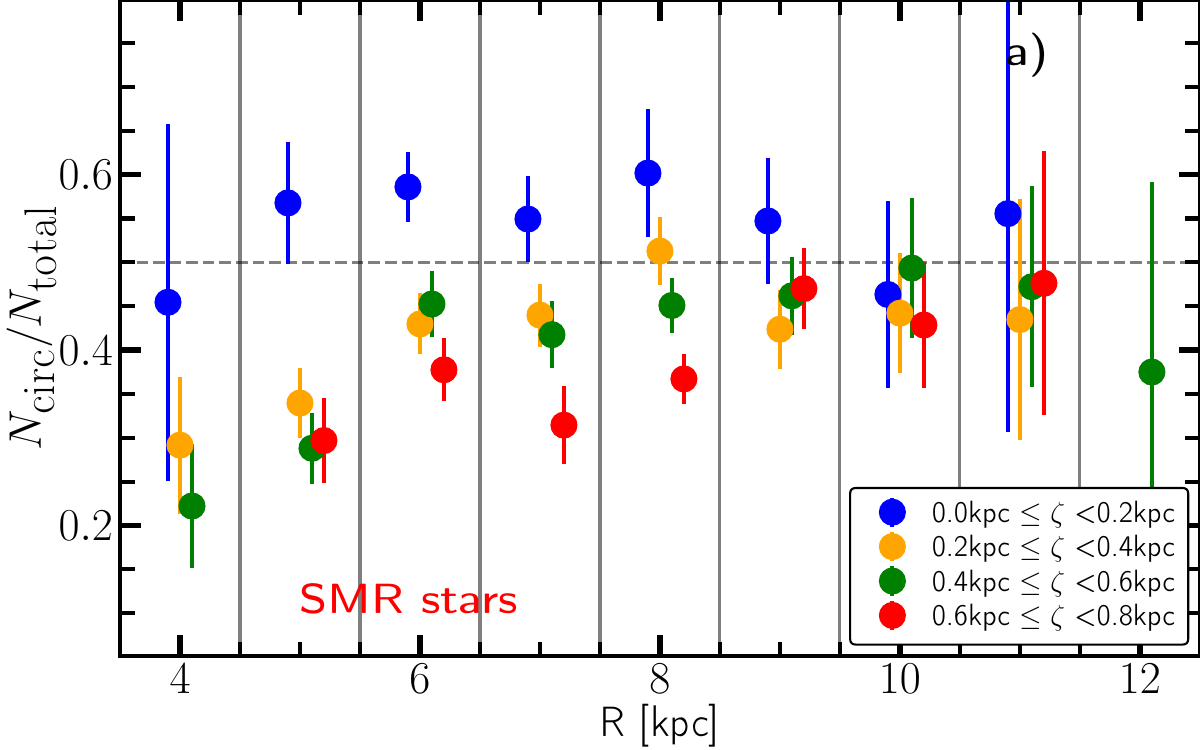}
    \includegraphics[width=0.47\textwidth]{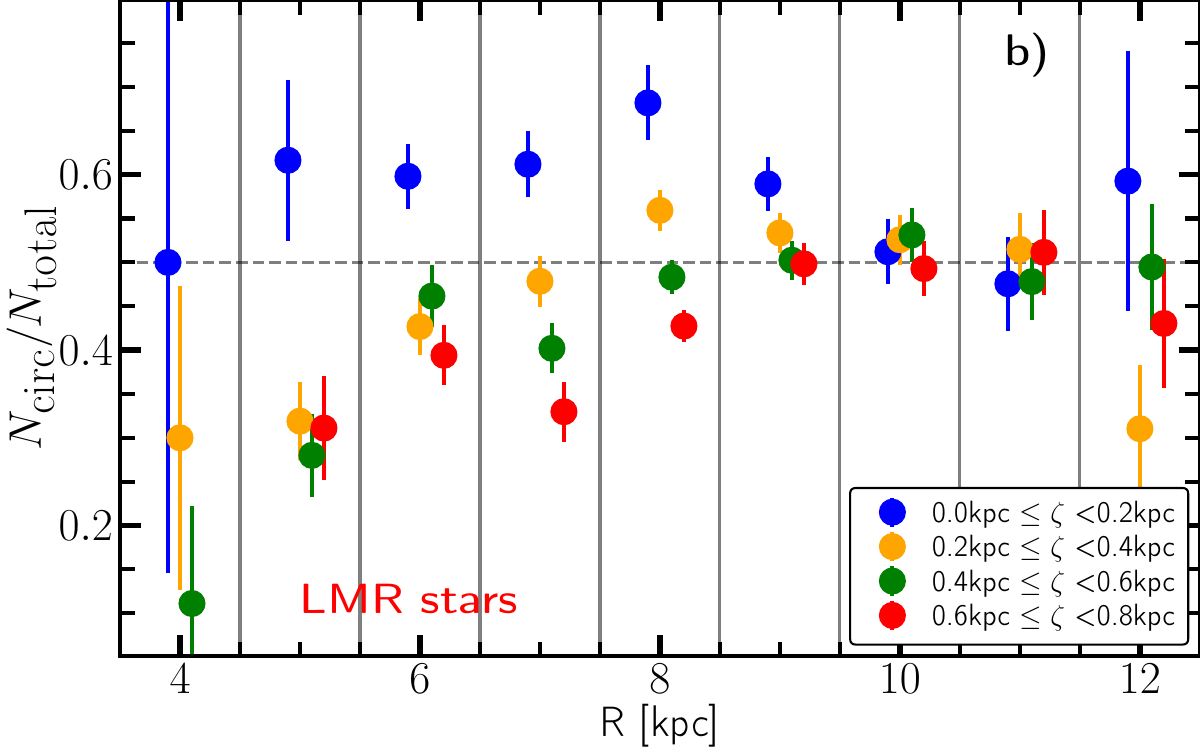}
    \caption{Fraction of stars on circular orbits relative to the total number of stars, split into $R$-bins. The SMR and LMR selections are further split into bins with different $\zeta$ (see \Aref{app:gauge} for a detailed explanation of $\zeta$).}
    \label{fig:ecc_R_kord}
\end{figure}

In \Fref{fig:Kord_comparison} and \Fref{fig:ecc_R_kord}, we explore the fraction of stars on circular orbits, which is used as a proxy to identify the primary migration mechanism as churning or blurring, in different bins of both $R$ and $\zeta$. Our $\zeta$ is a modified $z_\mathrm{max}$ scaled by the strength of the Galactic potential at a given $R$ (see \Aref{app:gauge} for details). In \Fref{fig:Kord_comparison} we compare our sample with the results from \citetalias{Kordopatis2015}, applying the same $|z|<\SI{400}{pc}$ selection. Our SMR definition is the most directly comparable to the results given in Table 3 of \citetalias{Kordopatis2015}. Specifically, we combined the columns with $\mathrm{[Fe/H]}>\SI{0.2}{dex}$ into a single data point for each of the three available $R$ ranges in Table 3 (with weighting by the number of stars, which is given in their Figure 10).
We see agreement in the $\SI{7.5}{kpc}\leq R <\SI{8.5}{kpc}$ bin between our SMR selection and the \citetalias{Kordopatis2015} sample. On the other hand, the $\SI{6.5}{kpc}\leq R <\SI{7.5}{kpc}$ and $\SI{8.5}{kpc}\leq R <\SI{9.5}{kpc}$ bins are more circular in our sample.
We also show the LMR results of our sample to showcase the differences to our SMR selection and the \citetalias{Kordopatis2015} data.

Furthermore, we explored trends of circularity with changing $\zeta$ in \Fref{fig:ecc_R_kord}\,a (for SMR stars) and \Fref{fig:ecc_R_kord}\,b (for LMR stars). We see that the fraction of stars on circular orbits changes as a function of $\zeta$, especially between $\SI{4.5}{kpc}\leq R <\SI{9.5}{kpc}$ where the lowest $\zeta$ sample stands out. These are the stars orbiting closest to the Galactic midplane. In the $\SI{7.5}{kpc}\leq R <\SI{9.5}{kpc}$ bins we see a trend of decreasing fractional circularity with increasing $\zeta$.
These trends are not present further outwards in the disc. Our sample is not large enough to explore the Galaxy at $R <\SI{4.5}{kpc}$.

\subsection{Eccentricity of the population}
\begin{figure*}
    \centering
    \includegraphics[width=0.9\textwidth]{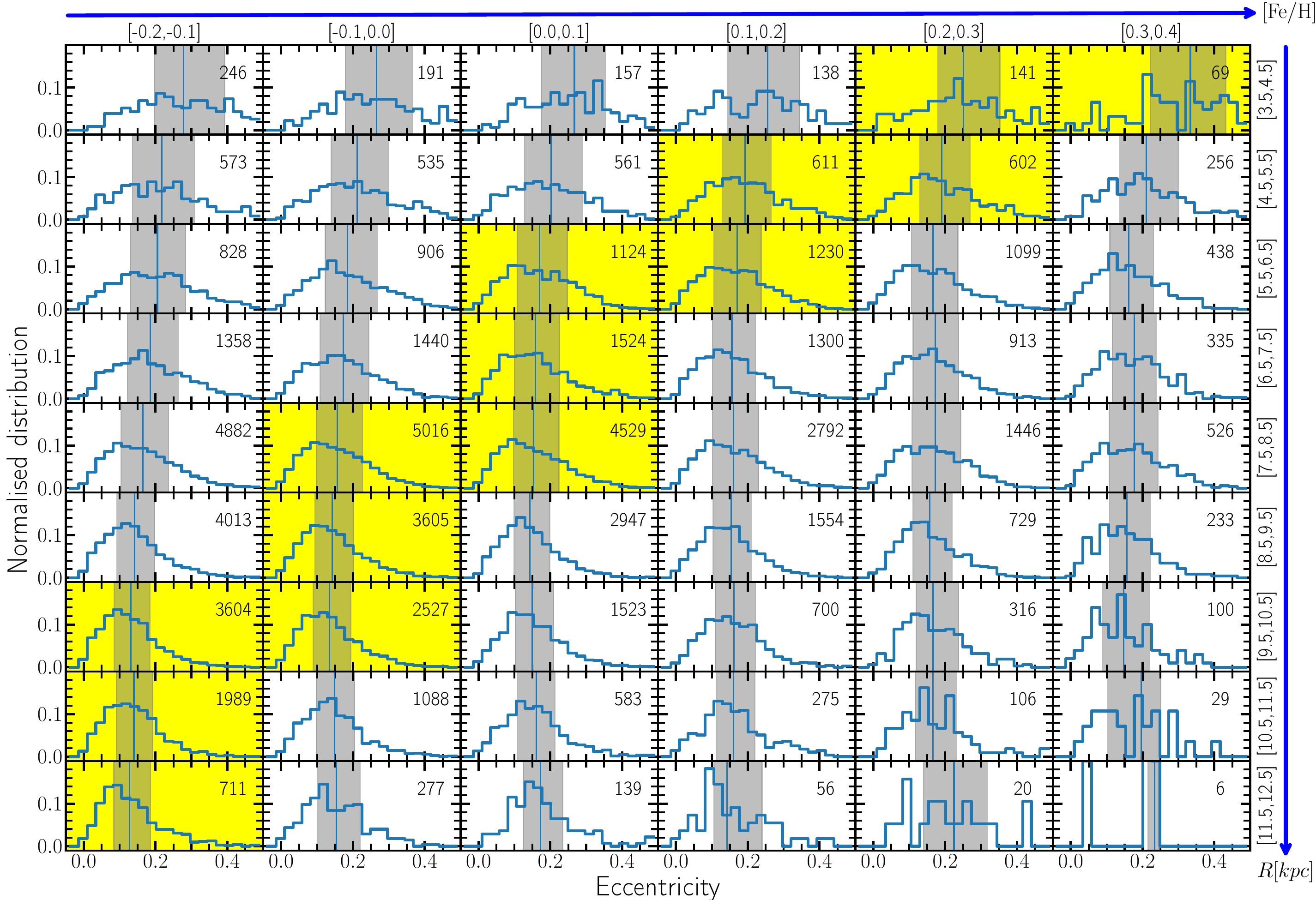}
    \caption{Eccentricity of stellar orbits for stars in our full sample split into bins of Galacto-centric radius $R$ and $\mathrm{[Fe/H]}$. The histograms are individually normalised to $1$. The panels become more metal-rich towards the right and increase in $R$ towards the bottom. The grey area represents 50\% of all stars within a bin, i.e. ranging between the 25th and 75th percentile, with the vertical line representing the median. We did not select in $z$ in this figure and the number of stars in each panel is shown on the top right. The panels indicated in yellow show the metallicity ranges that the current day ISM crosses within that panel's range of $R$. Multiple panels may be marked if the ISM crosses both of them within the $R$ range.}
    \label{fig:ecc_hist}
\end{figure*}

In \Fref{fig:ecc_hist} we show histograms of the orbital eccentricity for stars in bins of metallicity and Galactocentric radius. The number of stars is highest in the bins corresponding the the metallicity of the present day ISM at any given Galactocentric radius (with some exceptions). These [Fe/H]-$R$ bins are highlighted in yellow in \Fref{fig:ecc_hist}.
In the inner Galaxy ($R<\SI{6.5}{kpc}$), the eccentricity distribution of our sample does not change much with increasing metallicity. We note that, given our assumption that stars are born with a metallicity corresponding to the ISM metallicity of their local environment, all stars in panels to the right of the yellow panels should contain stars that must have migrated outwards.

\subsection{Age of the population}
\begin{figure*}
    \centering
    \includegraphics[width=0.9\textwidth]{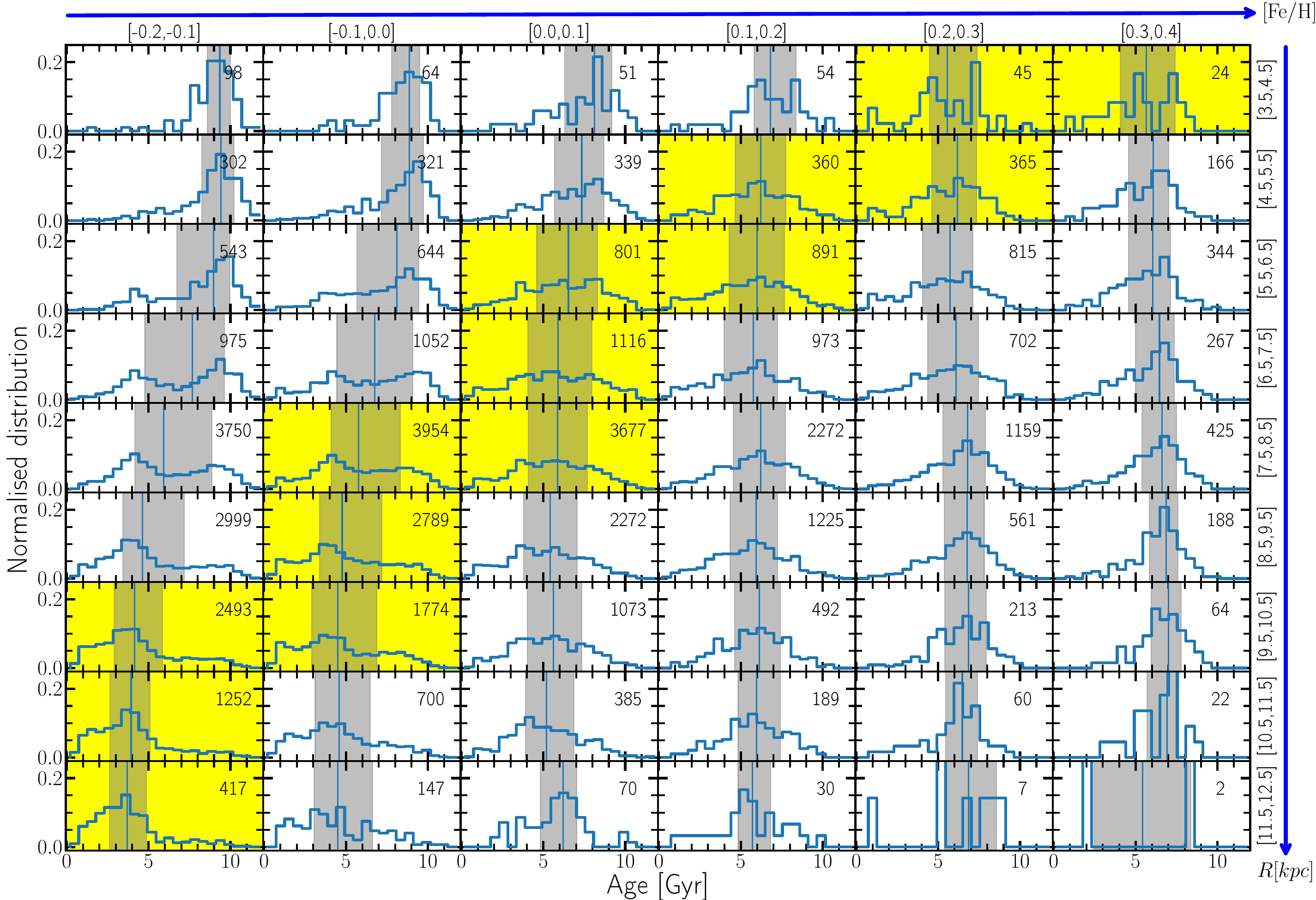}
    \caption{Ages of stars for our full sample (where ages were available) split into bins of Galactocentric radius $R$ and $\mathrm{[Fe/H]}$. The panels become more metal-rich towards the right and increase in $R$ towards the bottom. The grey area represents 50\% of all stars within a bin (from the 25th to 75th percentile) with the vertical line representing the median. We did not select in $z$ in this figure and the number of stars in each panel is shown on the top right. The yellow panels highlight each panel which contains the ISM metallicity somewhere within their range of Galactocentric radius.}
    \label{fig:age_hist}
\end{figure*}

In \Fref{fig:age_hist} we show histograms of the ages for stars in bins of metallicity and Galactocentric radius. The ages are taken from \citet{Anders2023}, who use XGBoost trained on asteroseismic ages to determine ages of APOGEE\,DR17 giant stars. These ages have a median relative statistical uncertainty of 17\%.
The median age of the metal-poor stars ($\mathrm{[Fe/H]}<0.0$) decreases with increasing Galactocentric radius, reaching a minimum at the radius where the  metallicity of the bin is equivalent to that of the local ISM. On the metal-rich end, the median stellar age increases with increasing Galactocentric radius.

\section{Discussion}
\label{sect:disc}
How effective both churning and blurring processes are at migrating stars outwards through the disc is a leading question of this work. The main issue we encountered is that churning does not influence the eccentricity of the orbit \citep{Sellwood2002}, i.e.\ the orbit of a churned star is indistinguishable from those of stars that formed at the new radius on a circular orbit. This means our main tracer for churning is the comparison of a star's metallicity to the local ISM, where we assume that a star's metallicity reflects the composition of the local ISM at formation \citep{Minchev2018, Minchev2020}.

Alternative explanations for the SMR stars presently found in the outer regions of the Milky Way disc include accretion from low-mass galaxies or in-situ formation from locally enriched material (so called Galactic fountains). An accreted origin is unlikely as observations of nearby satellite galaxies find these stars to have $\mathrm{[Fe/H]} \lessapprox 0$ \citep[][and references therein]{Tolstoy2009}. Metal-rich Galactic fountains \citep[][]{Chen2019} are unlikely to provide enough material to form SMR stars in the numbers we observe. In addition, the age trends of our sample, shown in \Fref{fig:age_hist}, suggest that the metallicity of the youngest stars generally follows the metallicity of the local ISM.

\subsection[Differences between our results and Kordopatis et al. (2015)]{Differences between our results and \citetalias{Kordopatis2015}}
In the top panel of \Fref{fig:ecc_R_kord}, we can see the results from \citetalias{Kordopatis2015} using RAVE data compared to our results using APOGEE data. We note that the metallicity scale of RAVE and APOGEE\,DR17 are very similar, see \citet{2022A&A...663A...4S}. Importantly, in the solar neighbourhood bin ($\SI{7.5}{kpc}<R<\SI{8.5}{kpc}$) our SMR selection agrees with the \citetalias{Kordopatis2015} results.
These two points probe the same volume and use the same metallicity selection.
The other radial bins in common ($\SI{6.5}{kpc}<R<\SI{7.5}{kpc}$ and $\SI{8.5}{kpc}<R<\SI{9.5}{kpc}$) are different between the studies. One explanation for this difference is the volume probed by the APOGEE and RAVE surveys, which would affect the most distant targets more severely. In the case of RAVE that would be red giants at a distance of up to $\SI{1.5}{kpc}$. The results of \citetalias{Kordopatis2015} are based on RAVE DR4 \citep{Kordopatis2013}, which means that at that time, \textit{Gaia} kinematic information was not available and that further out stars might have suffered from poor kinematic and distance measurements.

\subsection{Tracing the efficiency of churning and blurring}
We have shown that the efficiency of churning varies as a function of $\zeta$ and $z_\mathrm{max}$ (see \Sref{sec:chur_zmax} and \Aref{app:gauge}). About half of our sample using either selection shows at least some amount of blurring (i.e. non-circular orbits) at all $R$ and $\zeta$ (\Fref{fig:ecc_R_kord}).

Our LMR selection was explicitly designed to identify stars with metallicities that cannot be explained entirely by blurring. By using the guiding radius of the stellar orbit to compare with the radial metallicity gradient, we select only stars that have an orbital guiding radius outwards of their possible birth radius.
Therefore, to explain their metal enrichment we assume that all of the LMR stars have undergone some churning (with most stars showing signs of blurring as well). As seen in \Fref{fig:ecc_hist}, this is a large number of stars at any given $R$.

We conclude that most radial migration in our samples of LMR and SMR stars is caused by a combination of both churning and blurring. Our findings suggest that churning is more dominant in the Galactic midplane, while the influence of blurring increases with increasing $\zeta$ (or $z_\mathrm{max}$). 

\subsection[Churning sensitivity to zmax]{Churning sensitivity to $z$ and $z_\mathrm{max}$}\label{sec:chur_zmax}
\Fref{fig:ecc_R_kord} indicates that churning  becomes less efficient for stars that spend more time further away from the disc during their orbits.
In the work of \citetalias{Kordopatis2015}, a selection based on the vertical position was included, i.e.\ $|z|<\SI{400}{pc}$.
The biggest advantage of using  $z$ is that it is a simple quantity to understand and does not require extrapolation of the star's orbit, which can, if some of the data are poor, acquire large uncertainties.
However, using $z_\mathrm{max}$ provides a much better picture of the overall movement of the star, rather than a snapshot of a single moment in time. Especially when it comes to radial migration, trends in the dominant migration mechanism as a functions of orbital $z_\mathrm{max}$ can place constraints on the efficiency of churning (if our results from \Sref{sec:analysis} hold true). Therefore, we prefer to use our $\zeta$ parameter (which is a modified $z_\mathrm{max}$, see \Aref{app:gauge}) as the main quantity to explore potential trends in the vertical component of the stellar orbits.

We use the fraction of stars on circular orbits as a proxy for the efficiency of blurring compared to the efficiency of churning. However, there are two trends to keep in mind: (i) heating causes an increase in both eccentricity and $z_\mathrm{max}$, meaning that an orbit with higher $z_\mathrm{max}$ should naturally be less circular. However, we can still interpret the results from \Fref{fig:ecc_R_kord}, because we can compare the fraction of circular orbits relative to the total number of stars in our sample at different Galactocentric radius. (ii) Stars with a guiding radius close to $R_\odot=\SI{8.249}{kpc}$ are on average closer to us when on a circular orbit while the opposite is true for stars further away (at $R_G=\SI{3}{kpc}$ or $R_G=\SI{11}{kpc}$).
This means that naturally the number of stars on circular orbits will peak around solar Galactocentric radius (as seen in \Fref{fig:ecc_R_kord}). This should be taken into account when interpreting our results.

As seen in \Fref{fig:ecc_R_kord}, the efficiency of churning seems to be highest for stars with the lowest $\zeta$, while stars with progressively higher $\zeta$ have a declining fraction of circular orbits. The bins spanning $\SI{5.5}{kpc}<R<\SI{9.5}{kpc}$ show this clearly. This could mean that churning processes require the star to be close to the Galactic plane at the time of churning, which is more likely for stars with low $\zeta$.
\citet{Minchev2012} found in their simulation that the vertical (i.e. in $z$-direction) action is conserved when stars are churned, which seems to be in agreement with our own results. 
Our results towards the inner Galaxy ($\SI{5.5}{kpc}$ to $\SI{9.5}{kpc}$, see \Fref{fig:ecc_R_kord}) suggest that the blurring and/or heating processes in the inner Galaxy might not conserve the vertical action as their efficiency depends on $\zeta$. However, this does not hold true for the outer Galaxy ($>\SI{9.5}{kpc}$, see \Fref{fig:ecc_R_kord}), which means further investigation would be needed to understand this fully.

\subsection{Age trends}
\label{sec:age_trends}
Ages trend in our sample can be seen in \Fref{fig:R_FeH_age}b and \Fref{fig:age_hist}. In \Fref{fig:R_FeH_age}b we note that the $R_G$ - $\mathrm{[Fe/H]}$ bins with the youngest median age all lie metal-poor of the LMR selection line. This suggests most young stars have a metallicity similar to or below the local ISM metallicity.
In \Fref{fig:age_hist} we see that at all guiding radii, the youngest stars have [Fe/H] within $\SI{0.1}{dex}$ of the local ISM metallicity. Both figures confirm the idea that very young stars have not had time to significantly radial migrate, which is a similar conclusion obtained with a different sample \citet{ViscasillasVazquez2023} ($\sim8000$ stars in open clusters based on RVS data).

In \Fref{fig:R_FeH_age}, the presence of large numbers of stars more metal-rich than the local ISM line is significant, and the inspiration for this work. Interestingly, we see that the median age of the LMR sample is not young at all; at all $R_G$ and $\mathrm{[Fe/H]}$, the LMR sample has a median age that is $\sim4-\SI{8}{Gyr}$ and does not strongly vary with either $R_G$ or $\mathrm{[Fe/H]}$.

In both \Fref{fig:R_FeH_age} b and \Fref{fig:age_hist}, we see a strong trend in the median age of relatively metal-poor stars ($\SI{-0.2}{dex}\leq\mathrm{[Fe/H]}\leq\SI{0.0}{dex}$) such that a population of a given $\mathrm{[Fe/H]}$ becomes younger with increasing $R/R_G$ until the $R/R_G$ where the stellar $\mathrm{[Fe/H]}$ matches the ISM $\mathrm{[Fe/H]}$. 
Our interpretation of this trend is that at a given $R$, the locally formed stellar population dominates. Generally, the ISM is enriched over time, so at any $R$, the locally formed stars have correspondingly higher $\mathrm{[Fe/H]}$ with time. 
In the inner Galaxy, the metal-poor stars formed earlier (and are now old) and stars with equally low $\mathrm{[Fe/H]}$ do not form there anymore. The result is that, at a given $R$, in the $\mathrm{[Fe/H]}$ bins that approach the local ISM $\mathrm{[Fe/H]}$, the median age decreases because the dominant locally formed stars have formed more recently. 
Some panels in \Fref{fig:age_hist} show a peak at very young ages, which could indicates a recent infall of new star-forming material \citep{Spitoni2023}. Since this is not consistently present in panels of similar metallicity, we cannot make a definitive statement for or against this.

Metal-rich stars on the other hand must form in metal-rich environments, i.e. at low $R$. This means that stars that populate the panels more to the right and down from the ISM panels (yellow panels in \Fref{fig:age_hist}) cannot be stars that were formed from the local ISM. Since the ages of these populations seem to plateau as a function of $R$, we assume that most of these metal-rich stars belong to the same initial population of inner Galaxy metal-rich stars and the population further outward underwent radial migration to be observed at their current location.
This supports the idea of most stars were formed from material within the Milky Way with a strict radial metallicity gradient, rather than significant amounts of stars formed from metal-rich fountains \citep{Chen2019} or accretion from dwarf spheroidal galaxies, which at most are observed to reach solar metallicity \citep{Mucciarelli2017}.

\section{Conclusions}\label{sec:conclusion}
We analysed a subset of red giants from APOGEE DR17 to explore radial migration, focusing on the difference between migration via churning and blurring. Assuming stars form from the gas constituting the local ISM around them, we found that a large fraction of metal-rich stars experience some amount of churning. This explains the difference between their metallicity and the relatively low metallicity of their surrounding ISM. However, in most super metal-rich and local metal-rich populations (\Sref{sec:metal_rich_def}), we find large fractions of stars on eccentric orbits which indicates a significant amount of blurring in their stellar history. Therefore, we would like to propose the view that most stars migrate through a combination of both churning and blurring (with varying efficiency in individual stars). This is in line with what has been found in previous studies, \citet{Halle2015} which shows simulations in which migration could not be explained by blurring alone. Furthermore, \citet{Feltzing2020} found that about half of the stars that migrated from their formation radius (red giant branch stars from APOGEE DR14) experienced both churning and blurring, with a smaller fraction experiencing only one of these effects. Our LMR selection aims at stars that migrated via churning, so likely all of them migrated through churning, while the SMR selection consists of $>90\%$ LMR stars. With this prevalence of churning in these stars, we found that about 50\% of either selection of stars were additionally blurred enough to be on non-circular orbits (ecc>0.15).

We also found that the overall efficiency of churning and blurring compared to each other vary with $z_\mathrm{max}$ (and $\zeta$) in our sample, i.e. metal-rich stars with $z_\mathrm{max}<\SI{200}{pc}$ experience relatively more churning and stars with $\SI{200}{pc}<z_\mathrm{max}<\SI{800}{pc}$ experienced relatively more blurring.
This effectiveness might originate from the fact that churning via gravitational effects of the spiral arms is most effective for stars that are close to $z=0$ while blurring sources do not necessarily depend on $z_\mathrm{max}$. Another view of this would be that stars on eccentric orbits (i.e. stars with orbits that have been blurred) are more likely to be heated vertically as they encounter more structures for heating interactions. This might be cause for further investigation.

We compared our sample with the RAVE sample used in an earlier study \citep{Kordopatis2015}, and found agreement in the solar neighbourhood but disagreement when extending to bins further away in the radial direction. This could originate from a range of sources taking effect at larger distances ($>\SI{1}{kpc}$ from the Sun), e.g.\ limitations in the selection functions of stars at further distances or larger uncertainties/systematic errors on their kinematic measurements in the pre-\textit{Gaia} era.

Lastly, we investigate the age of our red giant population with regards to their radial position and metallicity. We found that the overall population follows a few trends we found significant, i.e. that metal-poor stars ($-0.2<\mathrm{[Fe/H]}<0.0$) are generally younger in the outer disc than in the inner disc while metal-rich stars ($\mathrm{[Fe/H]}>0.0$) in the outer disc have similar(or older) ages than inner disc stars.
This supports the idea that the majority of stars form from their local ISM as the youngest stars are following the ISM metallicity curve and then either migrate or stay locally, which is why we find old metal-poor populations further inwards in the disc where the ISM metallicity exceed that of these stars. 

In future works, it would be interesting to look at variations of migration patterns at different azimuthal angles in the Milky Way to relate the churning and blurring efficiency to the distance from larger structures (i.e. the spiral arms). We found that the APOGEE sample does not cover a sufficient range of azimuthal angle to investigate this. Roughly $71\%$ of the stars in our sample have $-15^\circ<\phi<15^\circ$ (where $\phi=0$ for the Sun) which is a too small angle given the total number of stars to make any division into angular ranges feasible. Within the available range of azimuthal angles in our selected sample, we found no evidence for variations of migratory behaviour.
We suggest that data with larger variations in azimuthal angle, such as potentially the 4MOST or WEAVE catalogues, will allow us to study additional migration patterns.

\section*{Acknowledgements}
We thank Nikos Prantzos for making available his age-metallicity curve to us.

C.L., S.F., and D.F. were supported by a project grant from the Knut and Alice Wallenberg Foundation (KAW 2020.0061 Galactic Time Machine, PI Feltzing). 
G.K. gratefully acknowledges support from the French National Research Agency (ANR) funded project “MWDisc” (ANR-20-CE31-0004). This project was also supported by funds from the Crafoord foundation (reference 20230890).

This work has made use of data from the European Space Agency (ESA) mission Gaia (\url{https://www.cosmos.esa.int/gaia}), processed by the Gaia Data Processing and Analysis Consortium (DPAC; \url{https://www.cosmos.esa.int/web/gaia/dpac/consortium}). Funding for the DPAC has been provided by national institutions, in particular the institutions participating in the Gaia Multilateral Agreement. Funding for the Sloan Digital Sky Survey IV has been provided by the Alfred P. Sloan Foundation, the U.S. Department of Energy Office of Science, and the Participating Institutions. SDSS-IV acknowledges support and resources from the Center for High Performance Computing at the University of Utah. The SDSS website is \url{www.sdss4.org}.

SDSS-IV is managed by the Astrophysical Research Consortium for the Participating Institutions of the SDSS Collaboration including the Brazilian Participation Group, the Carnegie Institution for Science, Carnegie Mellon University, Center for Astrophysics-Harvard \& Smithsonian, the Chilean Participation Group, the French Participation Group, Instituto de Astrof\'{i}sica de Canarias, The Johns Hopkins University, Kavli Institute for the Physics and Mathematics of the Universe (IPMU) / University of Tokyo, the Korean Participation Group, Lawrence Berkeley National Laboratory, Leibniz Institut f\"{u}r Astrophysik Potsdam (AIP), Max-Planck-Institut f\"{u}r Astronomie (MPIA Heidelberg), Max-Planck-Institut f\"{u}r Astrophysik (MPA Garching), Max-Planck-Institut f\"{u}r Extraterrestrische Physik (MPE), National Astronomical Observatories of China, New Mexico State University, New York University, University of Notre Dame, Observat\'{a}rio Nacional/MCTI, The Ohio State University, Pennsylvania State University, Shanghai Astronomical Observatory, United Kingdom Participation Group, Universidad Nacional Aut\'{o}noma de M\'{e}xico, University of Arizona, University of Colorado Boulder, University of Oxford, University of Portsmouth, University of Utah, University of Virginia,  University of Washington, University of Wisconsin, Vanderbilt University, and Yale University.

We make use of the python \citep{python} packages Astropy \citep{astropy13, astropy18, astropy22}, Galpy \citep{galpy15}, Matplotlib \citep[][]{matplotlib}, Numpy \citep[][]{numpy}, and CMasher \citep{Velden2020}.

\section*{Data Availability}
We made use of APOGEE DR17 \citep{Abdurrouf2022}, which can be accessed here: \href{https://www.sdss4.org/dr17/data_access/}{\url{https://www.sdss4.org/dr17/data_access/}}. Furthermore, we used age estimates from \citet{Anders2023}, which can be accessed at \href{https://cdsarc.cds.unistra.fr/viz-bin/cat/J/A+A/678/A158}{https://cdsarc.cds.unistra.fr/viz-bin/cat/J/A+A/678/A158}, and the measurements from Gaia DR3 accessed at \href{https://gea.esac.esa.int/archive/}{https://gea.esac.esa.int/archive/}.



\bibliographystyle{mnras}
\bibliography{Bibliography} 

\begin{thebibliography}{}
\makeatletter
\relax
\def\mn@urlcharsother{\let\do\@makeother \do\$\do\&\do\#\do\^\do\_\do\%\do\~}
\def\mn@doi{\begingroup\mn@urlcharsother \@ifnextchar [ {\mn@doi@}
  {\mn@doi@[]}}
\def\mn@doi@[#1]#2{\def\@tempa{#1}\ifx\@tempa\@empty \href
  {http://dx.doi.org/#2} {doi:#2}\else \href {http://dx.doi.org/#2} {#1}\fi
  \endgroup}
\def\mn@eprint#1#2{\mn@eprint@#1:#2::\@nil}
\def\mn@eprint@arXiv#1{\href {http://arxiv.org/abs/#1} {{\tt arXiv:#1}}}
\def\mn@eprint@dblp#1{\href {http://dblp.uni-trier.de/rec/bibtex/#1.xml}
  {dblp:#1}}
\def\mn@eprint@#1:#2:#3:#4\@nil{\def\@tempa {#1}\def\@tempb {#2}\def\@tempc
  {#3}\ifx \@tempc \@empty \let \@tempc \@tempb \let \@tempb \@tempa \fi \ifx
  \@tempb \@empty \def\@tempb {arXiv}\fi \@ifundefined
  {mn@eprint@\@tempb}{\@tempb:\@tempc}{\expandafter \expandafter \csname
  mn@eprint@\@tempb\endcsname \expandafter{\@tempc}}}

\bibitem[\protect\citeauthoryear{Abdurro'uf et~al.,}{Abdurro'uf
  et~al.}{2022}]{Abdurrouf2022}
Abdurro'uf et~al., 2022, \mn@doi [ApJS] {10.3847/1538-4365/ac4414}, 259, 35

\bibitem[\protect\citeauthoryear{Anders et~al.,}{Anders
  et~al.}{2023}]{Anders2023}
Anders F.,  et~al., 2023, \mn@doi [A\&A] {10.1051/0004-6361/202346666}, 678,
  A158

\bibitem[\protect\citeauthoryear{Arellano-C\'ordova, Esteban, Garc\'ia-Rojas
  \& M\'endez-Delgado}{Arellano-C\'ordova et~al.}{2020}]{ArellanoCordova2020}
Arellano-C\'ordova K.~Z.,  Esteban C.,  Garc\'ia-Rojas J.,   M\'endez-Delgado
  J.~E.,  2020, \mn@doi [MNRAS] {10.1093/mnras/staa1523}, 496, 1051

\bibitem[\protect\citeauthoryear{{Armillotta}, {Krumholz}  \&
  {Fujimoto}}{{Armillotta} et~al.}{2018}]{Armillotta2018}
{Armillotta} L.,  {Krumholz} M.~R.,   {Fujimoto} Y.,  2018, \mn@doi [MNRAS]
  {10.1093/mnras/sty2625}, \href
  {https://ui.adsabs.harvard.edu/abs/2018MNRAS.481.5000A} {481, 5000}

\bibitem[\protect\citeauthoryear{{Astropy Collaboration} et~al.,}{{Astropy
  Collaboration} et~al.}{2013}]{astropy13}
{Astropy Collaboration} et~al., 2013, \mn@doi [A\&A]
  {10.1051/0004-6361/201322068}, 558, A33

\bibitem[\protect\citeauthoryear{{Astropy Collaboration} et~al.,}{{Astropy
  Collaboration} et~al.}{2018}]{astropy18}
{Astropy Collaboration} et~al., 2018, AJ, 156, 123

\bibitem[\protect\citeauthoryear{{Astropy Collaboration} et~al.,}{{Astropy
  Collaboration} et~al.}{2022}]{astropy22}
{Astropy Collaboration} et~al., 2022, \mn@doi [ApJ] {10.3847/1538-4357/ac7c74},
  935, 167

\bibitem[\protect\citeauthoryear{Bailer-Jones, Rybizki, Fouesneau, Demleitner
  \& Andrae}{Bailer-Jones et~al.}{2021}]{BailerJones2021}
Bailer-Jones C. A.~L.,  Rybizki J.,  Fouesneau M.,  Demleitner M.,   Andrae R.,
   2021, \mn@doi [ApJ] {10.3847/1538-3881/abd806}, 161, 147

\bibitem[\protect\citeauthoryear{Binney}{Binney}{2012}]{Binney2012}
Binney J.,  2012, \mn@doi [MNRAS] {10.1111/j.1365-2966.2012.21757.x}, 426, 1324

\bibitem[\protect\citeauthoryear{Blanc-Vaziaga, Cayrel  \&
  Cayrel}{Blanc-Vaziaga et~al.}{1973}]{BlancVaziaga1973}
Blanc-Vaziaga M.~J.,  Cayrel G.,   Cayrel R.,  1973, \mn@doi [ApJ]
  {10.1086/152013}, 180, 871

\bibitem[\protect\citeauthoryear{Bland-Hawthorn \& Gerhard}{Bland-Hawthorn \&
  Gerhard}{2016}]{BlandHawthorn2016}
Bland-Hawthorn J.,  Gerhard O.,  2016, \mn@doi [ARA\&A]
  {10.1146/annurev-astro-081915-023441}, 54, 529

\bibitem[\protect\citeauthoryear{Bono, Marconi, Cassisi, Caputo, Gieren  \&
  Pietrzynski}{Bono et~al.}{2005}]{Bono2005}
Bono G.,  Marconi M.,  Cassisi S.,  Caputo F.,  Gieren W.,   Pietrzynski G.,
  2005, \mn@doi [ApJ] {10.1086/427744}, 621, 966

\bibitem[\protect\citeauthoryear{Bovy}{Bovy}{2015}]{galpy15}
Bovy J.,  2015, \mn@doi [ApJS] {10.1088/0067-0049/216/2/29}, 216, 29

\bibitem[\protect\citeauthoryear{Bovy \& Rix}{Bovy \& Rix}{2013}]{Bovy2013}
Bovy J.,  Rix H.-W.,  2013, \mn@doi [ApJ] {10.1088/0004-637X/779/2/115}, 779,
  115

\bibitem[\protect\citeauthoryear{Buck}{Buck}{2020}]{Buck2020}
Buck T.,  2020, \mn@doi [MNRAS] {10.1093/mnras/stz3289}, 491, 5435

\bibitem[\protect\citeauthoryear{{Cayrel de Strobel}, {Bentolila}, {Hauck}  \&
  {Curchod}}{{Cayrel de Strobel} et~al.}{1980}]{1980A&AS...41..405C}
{Cayrel de Strobel} G.,  {Bentolila} C.,  {Hauck} B.,   {Curchod} A.,  1980,
  A{\&}AS, \href {https://ui.adsabs.harvard.edu/abs/1980A&AS...41..405C} {41,
  405}

\bibitem[\protect\citeauthoryear{Chen, Zhao, Zhao, Liang, Wu, Jia, Tian  \&
  Liu}{Chen et~al.}{2019}]{Chen2019}
Chen Y.~Q.,  Zhao G.,  Zhao J.~K.,  Liang X.~L.,  Wu Y.~Q.,  Jia Y.~P.,  Tian
  H.,   Liu J.~M.,  2019, \mn@doi [ApJ] {10.3847/1538-3881/ab5283}, 158, 249

\bibitem[\protect\citeauthoryear{{Dantas} et~al.,}{{Dantas}
  et~al.}{2023}]{2023A&A...669A..96D}
{Dantas} M.~L.~L.,  et~al., 2023, \mn@doi [A{\&}A]
  {10.1051/0004-6361/202243667}, \href
  {https://ui.adsabs.harvard.edu/abs/2023A&A...669A..96D} {669, A96}

\bibitem[\protect\citeauthoryear{De~Simone, Wu  \& Tremaine}{De~Simone
  et~al.}{2004}]{DeSimone2004}
De~Simone R.,  Wu X.,   Tremaine S.,  2004, \mn@doi [MNRAS]
  {10.1111/j.1365-2966.2004.07675.x}, 350, 627

\bibitem[\protect\citeauthoryear{Dehnen}{Dehnen}{1998}]{Dehnen1998}
Dehnen W.,  1998, \mn@doi [AJ] {10.1086/300364}, 115, 2384

\bibitem[\protect\citeauthoryear{Dehnen}{Dehnen}{2000}]{Dehnen2000}
Dehnen W.,  2000, \mn@doi [ApJ] {10.1086/301226}, 119, 800

\bibitem[\protect\citeauthoryear{Di~Matteo, Haywood, Combes, Semelin  \&
  Snaith}{Di~Matteo et~al.}{2013}]{DiMatteo2013}
Di~Matteo P.,  Haywood M.,  Combes F.,  Semelin B.,   Snaith O.~N.,  2013,
  \mn@doi [A\&A] {10.1051/0004-6361/201220539}, 553, A102

\bibitem[\protect\citeauthoryear{Di~Matteo et~al.,}{Di~Matteo
  et~al.}{2015}]{DiMatteo2015}
Di~Matteo P.,  et~al., 2015, \mn@doi [A\&A] {10.1051/0004-6361/201424457}, 577,
  A1

\bibitem[\protect\citeauthoryear{Donor et~al.,}{Donor et~al.}{2018}]{Donor2018}
Donor J.,  et~al., 2018, \mn@doi [AJ] {10.3847/1538-3881/aad635}, 156, 142

\bibitem[\protect\citeauthoryear{Feltzing, Bowers  \& Agertz}{Feltzing
  et~al.}{2020}]{Feltzing2020}
Feltzing S.,  Bowers J.~B.,   Agertz O.,  2020, \mn@doi [MNRAS]
  {10.1093/mnras/staa340}, 493, 1419

\bibitem[\protect\citeauthoryear{{Feng} \& {Krumholz}}{{Feng} \&
  {Krumholz}}{2014}]{Feng2014}
{Feng} Y.,  {Krumholz} M.~R.,  2014, \mn@doi [\nat] {10.1038/nature13662},
  \href {https://ui.adsabs.harvard.edu/abs/2014Natur.513..523F} {513, 523}

\bibitem[\protect\citeauthoryear{Frankel, Rix, Ting, Ness  \& Hogg}{Frankel
  et~al.}{2018}]{Frankel2018}
Frankel N.,  Rix H.-W.,  Ting Y.-S.,  Ness M.,   Hogg D.~W.,  2018, \mn@doi
  [ApJ] {10.3847/1538-4357/aadba5}, 865, 96

\bibitem[\protect\citeauthoryear{Freeman \& Bland-Hawthorn}{Freeman \&
  Bland-Hawthorn}{2002}]{Freeman2002}
Freeman K.,  Bland-Hawthorn J.,  2002, \mn@doi [ARA\&A]
  {10.1146/annurev.astro.40.060401.093840}, 40, 487

\bibitem[\protect\citeauthoryear{Fujimoto, Inutsuka  \& Baba}{Fujimoto
  et~al.}{2023}]{Fujimoto2023}
Fujimoto Y.,  Inutsuka S.-i.,   Baba J.,  2023, \mn@doi [MNRAS]
  {10.1093/mnras/stad1612}, 523, 3049

\bibitem[\protect\citeauthoryear{{GRAVITY Collaboration} et~al.,}{{GRAVITY
  Collaboration} et~al.}{2020}]{GRAVITYCollaboration2020}
{GRAVITY Collaboration} et~al., 2020, \mn@doi [A\&A]
  {10.1051/0004-6361/202037813}, 636, L5

\bibitem[\protect\citeauthoryear{{Gaia Collaboration} et~al.,}{{Gaia
  Collaboration} et~al.}{2023}]{2023A&A...674A...1G}
{Gaia Collaboration} et~al., 2023, \mn@doi [A{\&}A]
  {10.1051/0004-6361/202243940}, \href
  {https://ui.adsabs.harvard.edu/abs/2023A&A...674A...1G} {674, A1}

\bibitem[\protect\citeauthoryear{{Garc{\'\i}a P{\'e}rez} et~al.,}{{Garc{\'\i}a
  P{\'e}rez} et~al.}{2016}]{2016AJ....151..144G}
{Garc{\'\i}a P{\'e}rez} A.~E.,  et~al., 2016, \mn@doi [AJ]
  {10.3847/0004-6256/151/6/144}, \href
  {https://ui.adsabs.harvard.edu/abs/2016AJ....151..144G} {151, 144}

\bibitem[\protect\citeauthoryear{Grand et~al.,}{Grand et~al.}{2016}]{Grand2016}
Grand R. J.~J.,  et~al., 2016, \mn@doi [MNRAS] {10.1093/mnrasl/slw086}, 460,
  L94

\bibitem[\protect\citeauthoryear{Grenon}{Grenon}{1972}]{Grenon1972}
Grenon M.,  1972, in {Cayrel de Strobel} G.,  {Delplace} A.~M.,  eds, IAU
  Colloq. 17: Age des Etoiles. p.~55, \url
  {https://ui.adsabs.harvard.edu/abs/1972ade..coll...55G}

\bibitem[\protect\citeauthoryear{{Grenon}}{{Grenon}}{1999}]{1999Ap&SS.265..331G}
{Grenon} M.,  1999, \mn@doi [A{\&}AS] {10.1023/A:1002128300025}, \href
  {https://ui.adsabs.harvard.edu/abs/1999Ap&SS.265..331G} {265, 331}

\bibitem[\protect\citeauthoryear{Gustafsson, Church, Davies  \&
  Rickman}{Gustafsson et~al.}{2016}]{Gustafsson2016}
Gustafsson B.,  Church R.~P.,  Davies M.~B.,   Rickman H.,  2016, \mn@doi
  [A\&A] {10.1051/0004-6361/201423916}, 593, A85

\bibitem[\protect\citeauthoryear{Halle, Di~Matteo, Haywood  \& Combes}{Halle
  et~al.}{2015}]{Halle2015}
Halle A.,  Di~Matteo P.,  Haywood M.,   Combes F.,  2015, \mn@doi [A{\&}A]
  {10.1051/0004-6361/201525612}, 578, A58

\bibitem[\protect\citeauthoryear{H{\"a}nninen \& Flynn}{H{\"a}nninen \&
  Flynn}{2002}]{Haenninen2002}
H{\"a}nninen J.,  Flynn C.,  2002, \mn@doi [MNRAS]
  {10.1046/j.1365-8711.2002.05956.x}, 337, 731

\bibitem[\protect\citeauthoryear{Harris et~al.,}{Harris et~al.}{2020}]{numpy}
Harris C.~R.,  et~al., 2020, \mn@doi [Nature] {10.1038/s41586-020-2649-2}, 585,
  357

\bibitem[\protect\citeauthoryear{{Holtzman} et~al.,}{{Holtzman}
  et~al.}{2018}]{2018AJ....156..125H}
{Holtzman} J.~A.,  et~al., 2018, \mn@doi [AJ] {10.3847/1538-3881/aad4f9}, \href
  {https://ui.adsabs.harvard.edu/abs/2018AJ....156..125H} {156, 125}

\bibitem[\protect\citeauthoryear{Hunter}{Hunter}{2007}]{matplotlib}
Hunter J.~D.,  2007, Comput. Sci. Eng, 9, 90

\bibitem[\protect\citeauthoryear{{J{\"o}nsson} et~al.,}{{J{\"o}nsson}
  et~al.}{2018}]{2018AJ....156..126J}
{J{\"o}nsson} H.,  et~al., 2018, \mn@doi [AJ] {10.3847/1538-3881/aad4f5}, \href
  {https://ui.adsabs.harvard.edu/abs/2018AJ....156..126J} {156, 126}

\bibitem[\protect\citeauthoryear{Khoperskov, Di~Matteo, Haywood, G{\'o}mez  \&
  Snaith}{Khoperskov et~al.}{2020}]{Khoperskov2020}
Khoperskov S.,  Di~Matteo P.,  Haywood M.,  G{\'o}mez A.,   Snaith O.~N.,
  2020, \mn@doi [A\&A] {10.1051/0004-6361/201937188}, 638, A144

\bibitem[\protect\citeauthoryear{Kordopatis et~al.,}{Kordopatis
  et~al.}{2013}]{Kordopatis2013}
Kordopatis G.,  et~al., 2013, \mn@doi [ApJ] {10.1088/0004-6256/146/5/134}, 146,
  134

\bibitem[\protect\citeauthoryear{Kordopatis et~al.,}{Kordopatis
  et~al.}{2015}]{Kordopatis2015}
Kordopatis G.,  et~al., 2015, \mn@doi [MNRAS] {10.1093/mnras/stu2726}, 447,
  3526

\bibitem[\protect\citeauthoryear{Kordopatis et~al.,}{Kordopatis
  et~al.}{2023}]{Kordopatis2023}
Kordopatis G.,  et~al., 2023, \mn@doi [A\&A] {10.1051/0004-6361/202244283},
  669, A104

\bibitem[\protect\citeauthoryear{Kubryk, Prantzos  \& Athanassoula}{Kubryk
  et~al.}{2015}]{Kubryk2015}
Kubryk M.,  Prantzos N.,   Athanassoula E.,  2015, \mn@doi [A\&A]
  {10.1051/0004-6361/201424171}, 580, A126

\bibitem[\protect\citeauthoryear{Lu, Ness, Buck  \& Carr}{Lu
  et~al.}{2022}]{Lu2022}
Lu Y.~L.,  Ness M.~K.,  Buck T.,   Carr C.,  2022, \mn@doi [MNRAS]
  {10.1093/mnras/stac780}, 512, 4697

\bibitem[\protect\citeauthoryear{Mackereth et~al.,}{Mackereth
  et~al.}{2017}]{Mackereth2017}
Mackereth J.~T.,  et~al., 2017, \mn@doi [MNRAS] {10.1093/mnras/stx1774}, 471,
  3057

\bibitem[\protect\citeauthoryear{Mackereth et~al.,}{Mackereth
  et~al.}{2019}]{Mackereth2019}
Mackereth J.~T.,  et~al., 2019, \mn@doi [MNRAS] {10.1093/mnras/stz1521}, 489,
  176

\bibitem[\protect\citeauthoryear{Magrini et~al.,}{Magrini
  et~al.}{2023}]{Magrini2023}
Magrini L.,  et~al., 2023, \mn@doi [A\&A] {10.1051/0004-6361/202244957}, 669,
  A119

\bibitem[\protect\citeauthoryear{{Majewski} et~al.,}{{Majewski}
  et~al.}{2017}]{2017AJ....154...94M}
{Majewski} S.~R.,  et~al., 2017, \mn@doi [AJ] {10.3847/1538-3881/aa784d}, \href
  {https://ui.adsabs.harvard.edu/abs/2017AJ....154...94M} {154, 94}

\bibitem[\protect\citeauthoryear{McMillan}{McMillan}{2017}]{McMillan2017}
McMillan P.~J.,  2017, \mn@doi [MNRAS] {10.1093/mnras/stw2759}, 465, 76

\bibitem[\protect\citeauthoryear{Miglio et~al.,}{Miglio
  et~al.}{2021}]{Miglio2021}
Miglio A.,  et~al., 2021, \mn@doi [A\&A] {10.1051/0004-6361/202038307}, 645,
  A85

\bibitem[\protect\citeauthoryear{Minchev \& Quillen}{Minchev \&
  Quillen}{2006}]{Minchev2006}
Minchev I.,  Quillen A.~C.,  2006, \mn@doi [MNRAS]
  {10.1111/j.1365-2966.2006.10129.x}, 368, 623

\bibitem[\protect\citeauthoryear{Minchev, Famaey, Quillen, Dehnen, Martig  \&
  Siebert}{Minchev et~al.}{2012}]{Minchev2012}
Minchev I.,  Famaey B.,  Quillen A.~C.,  Dehnen W.,  Martig M.,   Siebert A.,
  2012, \mn@doi [A{\&}A] {10.1051/0004-6361/201219714}, 548, A127

\bibitem[\protect\citeauthoryear{Minchev et~al.,}{Minchev
  et~al.}{2018}]{Minchev2018}
Minchev I.,  et~al., 2018, \mn@doi [MNRAS] {10.1093/mnras/sty2033}, 481, 1645

\bibitem[\protect\citeauthoryear{Minchev, Anders  \& Chiappini}{Minchev
  et~al.}{2020}]{Minchev2020}
Minchev I.,  Anders F.,   Chiappini C.,  2020, in IAU General Assembly. pp
  253--254, \mn@doi{10.1017/S1743921319004216}, \url
  {https://ui.adsabs.harvard.edu/abs/2020IAUGA..30..253M}

\bibitem[\protect\citeauthoryear{Mucciarelli, Bellazzini, Ibata, Romano,
  Chapman  \& Monaco}{Mucciarelli et~al.}{2017}]{Mucciarelli2017}
Mucciarelli A.,  Bellazzini M.,  Ibata R.,  Romano D.,  Chapman S.~C.,   Monaco
  L.,  2017, \mn@doi [A{\&}A] {10.1051/0004-6361/201730707}, 605, A46

\bibitem[\protect\citeauthoryear{Myers et~al.,}{Myers et~al.}{2022}]{Myers2022}
Myers N.,  et~al., 2022, \mn@doi [AJ] {10.3847/1538-3881/ac7ce5}, 164, 85

\bibitem[\protect\citeauthoryear{Naiman et~al.,}{Naiman
  et~al.}{2018}]{Naiman2018}
Naiman J.~P.,  et~al., 2018, \mn@doi [MNRAS] {10.1093/mnras/sty618}, 477, 1206

\bibitem[\protect\citeauthoryear{Nair \& Abraham}{Nair \&
  Abraham}{2010}]{Nair2010}
Nair P.~B.,  Abraham R.~G.,  2010, \mn@doi [ApJS]
  {10.1088/0067-0049/186/2/427}, 186, 427

\bibitem[\protect\citeauthoryear{Nepal et~al.,}{Nepal et~al.}{2024}]{Nepal2024}
Nepal S.,  et~al., 2024, \mn@doi [A\&A] {10.1051/0004-6361/202348365}, 681, L8

\bibitem[\protect\citeauthoryear{Pietrukowicz, Soszy{\'n}ski  \&
  Udalski}{Pietrukowicz et~al.}{2021}]{Pietrukowicz2021}
Pietrukowicz P.,  Soszy{\'n}ski I.,   Udalski A.,  2021, \mn@doi [Acta
  Astronaut.] {10.32023/0001-5237/71.3.2}, 71, 205

\bibitem[\protect\citeauthoryear{Prantzos et~al.,}{Prantzos
  et~al.}{2023}]{Prantzos2023}
Prantzos N.,  et~al., 2023, \mn@doi [MNRAS] {10.1093/mnras/stad1551}, 523, 2126

\bibitem[\protect\citeauthoryear{Reid \& Brunthaler}{Reid \&
  Brunthaler}{2020}]{Reid2020}
Reid M.~J.,  Brunthaler A.,  2020, \mn@doi [ApJ] {10.3847/1538-4357/ab76cd},
  892, 39

\bibitem[\protect\citeauthoryear{Ro{\v{s}}kar, Debattista, Quinn, Stinson  \&
  Wadsley}{Ro{\v{s}}kar et~al.}{2008}]{Roskar2008}
Ro{\v{s}}kar R.,  Debattista V.~P.,  Quinn T.~R.,  Stinson G.~S.,   Wadsley J.,
   2008, \mn@doi [ApJL] {10.1086/592231}, 684, L79

\bibitem[\protect\citeauthoryear{Ro{\v{s}}kar, Debattista  \&
  Loebman}{Ro{\v{s}}kar et~al.}{2013}]{Roskar2013}
Ro{\v{s}}kar R.,  Debattista V.~P.,   Loebman S.~R.,  2013, \mn@doi [MNRAS]
  {10.1093/mnras/stt788}, 433, 976

\bibitem[\protect\citeauthoryear{Sch{\"o}nrich \& Binney}{Sch{\"o}nrich \&
  Binney}{2009}]{Schoenrich2009a}
Sch{\"o}nrich R.,  Binney J.,  2009, \mn@doi [MNRAS]
  {10.1111/j.1365-2966.2009.14750.x}, 396, 203

\bibitem[\protect\citeauthoryear{Sellwood}{Sellwood}{2014}]{Sellwood2014}
Sellwood J.~A.,  2014, \mn@doi [RMP] {10.1103/revmodphys.86.1}, 86, 1

\bibitem[\protect\citeauthoryear{Sellwood \& Binney}{Sellwood \&
  Binney}{2002}]{Sellwood2002}
Sellwood J.~A.,  Binney J.~J.,  2002, \mn@doi [MNRAS]
  {10.1046/j.1365-8711.2002.05806.x}, 336, 785

\bibitem[\protect\citeauthoryear{Sellwood \& Carlberg}{Sellwood \&
  Carlberg}{1984}]{Sellwood1984}
Sellwood J.~A.,  Carlberg R.~G.,  1984, \mn@doi [ApJ] {10.1086/162176}, 282, 61

\bibitem[\protect\citeauthoryear{Sellwood \& Masters}{Sellwood \&
  Masters}{2022}]{Sellwood2022}
Sellwood J.~A.,  Masters K.~L.,  2022, \mn@doi [ARA\&A]
  {10.1146/annurev-astro-052920-104505}, 60

\bibitem[\protect\citeauthoryear{Sharma et~al.,}{Sharma
  et~al.}{2021a}]{Sharma2021a}
Sharma S.,  et~al., 2021a, \mn@doi [MNRAS] {10.1093/mnras/stab1086}, 506, 1761

\bibitem[\protect\citeauthoryear{Sharma et~al.,}{Sharma
  et~al.}{2021b}]{Sharma2021}
Sharma S.,  et~al., 2021b, \mn@doi [MNRAS] {10.1093/mnras/stab3341}, 510, 734

\bibitem[\protect\citeauthoryear{{Soubiran}, {Le Campion}, {Cayrel de Strobel}
  \& {Caillo}}{{Soubiran} et~al.}{2010}]{2010A&A...515A.111S}
{Soubiran} C.,  {Le Campion} J.~F.,  {Cayrel de Strobel} G.,   {Caillo} A.,
  2010, \mn@doi [A{\&}A] {10.1051/0004-6361/201014247}, \href
  {https://ui.adsabs.harvard.edu/abs/2010A&A...515A.111S} {515, A111}

\bibitem[\protect\citeauthoryear{{Soubiran}, {Le Campion}, {Brouillet}  \&
  {Chemin}}{{Soubiran} et~al.}{2016}]{2016A&A...591A.118S}
{Soubiran} C.,  {Le Campion} J.-F.,  {Brouillet} N.,   {Chemin} L.,  2016,
  \mn@doi [A{\&}A] {10.1051/0004-6361/201628497}, \href
  {https://ui.adsabs.harvard.edu/abs/2016A&A...591A.118S} {591, A118}

\bibitem[\protect\citeauthoryear{{Soubiran}, {Brouillet}  \&
  {Casamiquela}}{{Soubiran} et~al.}{2022}]{2022A&A...663A...4S}
{Soubiran} C.,  {Brouillet} N.,   {Casamiquela} L.,  2022, \mn@doi [A\&A]
  {10.1051/0004-6361/202142409}, \href
  {https://ui.adsabs.harvard.edu/abs/2022A&A...663A...4S} {663, A4}

\bibitem[\protect\citeauthoryear{{Spinrad} \& {Taylor}}{{Spinrad} \&
  {Taylor}}{1969}]{1969ApJ...157.1279S}
{Spinrad} H.,  {Taylor} B.~J.,  1969, \mn@doi [ApJ] {10.1086/150154}, \href
  {https://ui.adsabs.harvard.edu/abs/1969ApJ...157.1279S} {157, 1279}

\bibitem[\protect\citeauthoryear{Spitoni et~al.,}{Spitoni
  et~al.}{2023}]{Spitoni2023}
Spitoni E.,  et~al., 2023, \mn@doi [A{\&}A] {10.1051/0004-6361/202244349}, 670,
  A109

\bibitem[\protect\citeauthoryear{Spitzer \& Schwarzschild}{Spitzer \&
  Schwarzschild}{1951}]{Spitzer1951}
Spitzer Lyman J.,  Schwarzschild M.,  1951, \mn@doi [ApJ] {10.1086/145478},
  114, 385

\bibitem[\protect\citeauthoryear{Tacconi, Genzel  \& Sternberg}{Tacconi
  et~al.}{2020}]{Tacconi2020}
Tacconi L.~J.,  Genzel R.,   Sternberg A.,  2020, \mn@doi [A\&AR]
  {10.1146/annurev-astro-082812-141034}, 58, 157

\bibitem[\protect\citeauthoryear{{Taylor}}{{Taylor}}{1996}]{1996ApJS..102..105T}
{Taylor} B.~J.,  1996, \mn@doi [ApJS] {10.1086/192253}, \href
  {https://ui.adsabs.harvard.edu/abs/1996ApJS..102..105T} {102, 105}

\bibitem[\protect\citeauthoryear{Tolstoy, Hill  \& Tosi}{Tolstoy
  et~al.}{2009}]{Tolstoy2009}
Tolstoy E.,  Hill V.,   Tosi M.,  2009, \mn@doi [ARA{\&}A]
  {10.1146/annurev-astro-082708-101650}, 47, 371

\bibitem[\protect\citeauthoryear{Van~Rossum \& Drake~Jr}{Van~Rossum \&
  Drake~Jr}{1995}]{python}
Van~Rossum G.,  Drake~Jr F.~L.,  1995, Python reference manual.
Centrum voor Wiskunde en Informatica Amsterdam

\bibitem[\protect\citeauthoryear{Viscasillas~V{\'a}zquez, Magrini, Spina,
  Tautvai{\v{s}}ien{\.{e}}, Van~der Swaelmen, Randich  \&
  Sacco}{Viscasillas~V{\'a}zquez et~al.}{2023}]{ViscasillasVazquez2023}
Viscasillas~V{\'a}zquez C.,  Magrini L.,  Spina L.,  Tautvai{\v{s}}ien{\.{e}}
  G.,  Van~der Swaelmen M.,  Randich S.,   Sacco G.~G.,  2023, \mn@doi [A{\&}A]
  {10.1051/0004-6361/202346963}, 679, A122

\bibitem[\protect\citeauthoryear{Vogelsberger, Marinacci, Torrey  \&
  Puchwein}{Vogelsberger et~al.}{2020}]{Vogelsberger2020}
Vogelsberger M.,  Marinacci F.,  Torrey P.,   Puchwein E.,  2020, \mn@doi [Nat.
  Rev. Phys.] {10.1038/s42254-019-0127-2}, 2, 42

\bibitem[\protect\citeauthoryear{Wenger, Balser, Anderson  \& Bania}{Wenger
  et~al.}{2019}]{Wenger2019}
Wenger T.~V.,  Balser D.~S.,  Anderson L.~D.,   Bania T.~M.,  2019, \mn@doi
  [ApJ] {10.3847/1538-4357/ab53d3}, 887, 114

\bibitem[\protect\citeauthoryear{Wielen, Fuchs  \& Dettbarn}{Wielen
  et~al.}{1996}]{Wielen1996}
Wielen R.,  Fuchs B.,   Dettbarn C.,  1996, A\&A, 314, 438

\bibitem[\protect\citeauthoryear{{Wilson} et~al.,}{{Wilson}
  et~al.}{2019}]{2019PASP..131e5001W}
{Wilson} J.~C.,  et~al., 2019, \mn@doi [\pasp] {10.1088/1538-3873/ab0075},
  \href {https://ui.adsabs.harvard.edu/abs/2019PASP..131e5001W} {131, 055001}

\bibitem[\protect\citeauthoryear{da Silva et~al.,}{da~Silva
  et~al.}{2023}]{Silva2023}
da Silva R.,  et~al., 2023, \mn@doi [A\&A] {10.1051/0004-6361/202346982}, 678,
  A195

\bibitem[\protect\citeauthoryear{van~der Velden}{van~der
  Velden}{2020}]{Velden2020}
van~der Velden E.,  2020, \mn@doi [JOSS] {10.21105/joss.02004}, 5, 2004

\makeatother
\end{thebibliography}




\appendix


\section{Inspection of spectra}
\label{app:spectra}

\begin{figure*}
    \centering
    \includegraphics[width=0.9\textwidth]{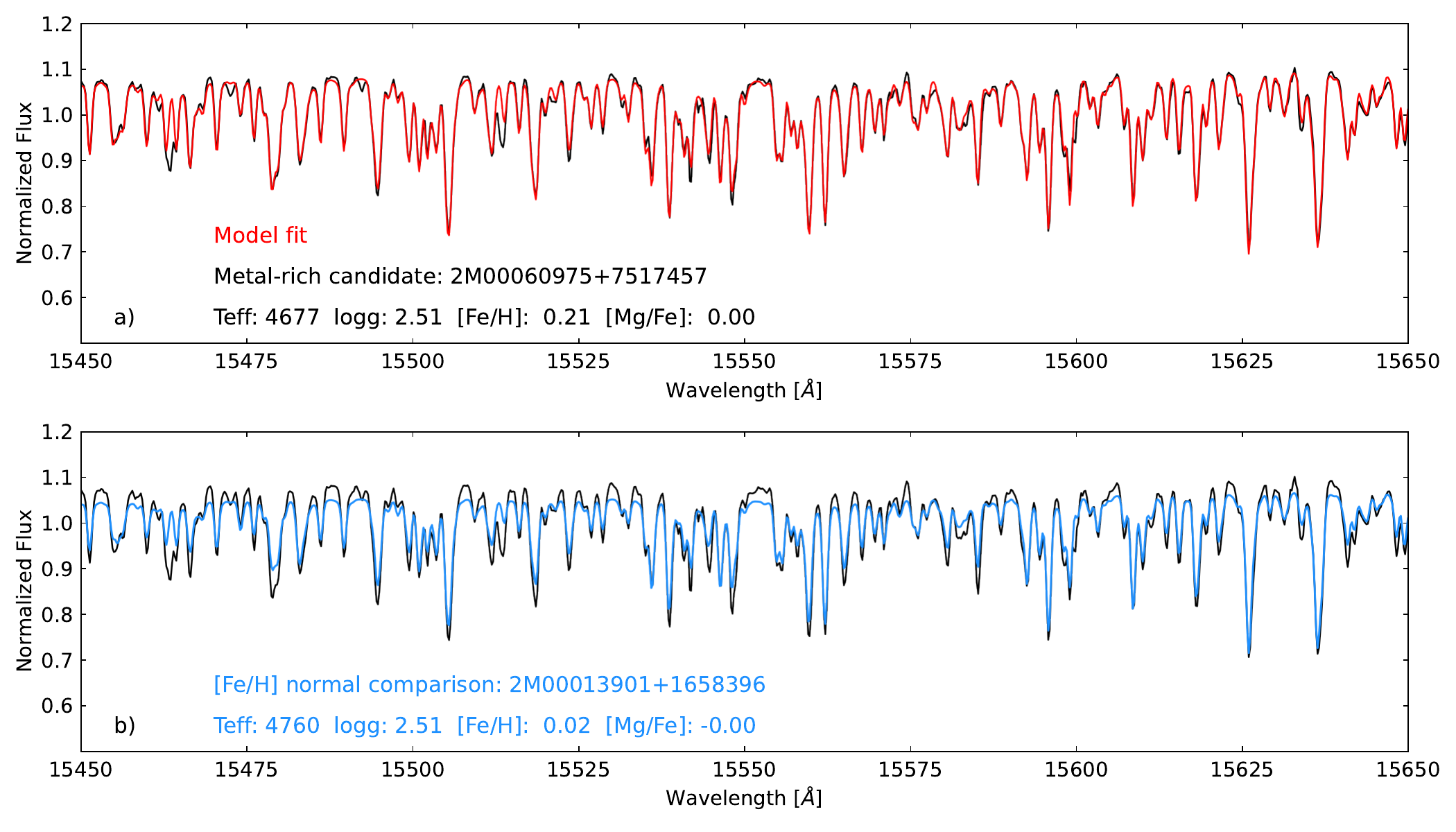}
    \caption{Example spectrum of a metal-rich star in our sample. The APOGEE {\tt aspcapStar} spectrum of 2M00060975+7517457 (black) compared with the synthetic model (red, panel a) and a solar metallicity star at the same $\log g$ (blue, panel b). The APOGEE derived T$_{\rm eff}$, $\log g$, $\mathrm{[Fe/H]}$, and $\mathrm{[Mg/Fe]}$ values of the metal-rich star and solar metallicity star are given.}
    \label{fig:checkspec}
\end{figure*}

To ensure the validity of the high $\mathrm{[Fe/H]}$ measurements, we have visually inspected spectra for a representative sub-sample of $\sim50$ the the most metal-rich stars covering a range of $\log g$ values. An example spectrum is shown in Fig. \ref{fig:checkspec} for the wavelength region $\SI{15450}{}$ to $\SI{15650}{\angstrom}$. For each star that we checked, we inspect the 1D normalised spectrum (black line) used by APOGEE to determine the stellar parameters and elemental abundances ({\tt aspcapStar} file) and how it compares to the synthetic model spectrum (red line, panel a). The final determined stellar parameters for the metal-rich star given in panel a. Generally, the model spectra fit the observed spectra well. We note that the red synthetic spectrum is not used to calculate the $\mathrm{[Fe/H]}$ abundance reported by APOGEE, but does represent the overall quality of the spectral analysis. The $\mathrm{[Fe/H]}$ abundance is calculated by fitting smaller spectral windows around individual Fe lines \citep[for more details see][]{2016AJ....151..144G, 2018AJ....156..125H, 2018AJ....156..126J}.

We also compare the spectrum of each metal-rich star to that of a star with similar $\log g$, but $\mathrm{[Fe/H]}\sim 0$ (blue line, panel b). The solar metallicity spectrum generally has weaker absorption lines, supporting the measurement of a lower metallicity. 

We find no evidence that the APOGEE measurements of $\mathrm{[Fe/H]}>~0.2$ are systematically overestimated compared with stars at solar metallicity with the same $\log g$.

\section[zmax dependence on the potential]{$z_\mathrm{max}$ dependence on the potential}\label{app:gauge}
In this work, the division of sub-samples via $z_\mathrm{max}$ has the purpose of identifying individual stars as part of a population of stars with specific energies in the $z$ direction. Since the potential of the Galactic disc varies in the radial direction through the disc, if we want to compare the motions of stars in the inner disc to stars in the outer disc as a function of the $z$ component of their orbit, we need to consider how the potential changes. We assume the Milky Way potential model of \citet{McMillan2017}, to account for the difference in the potential as a function of Galactocentric radius. 

\begin{figure}
    \centering
    \includegraphics[width=0.47\textwidth]{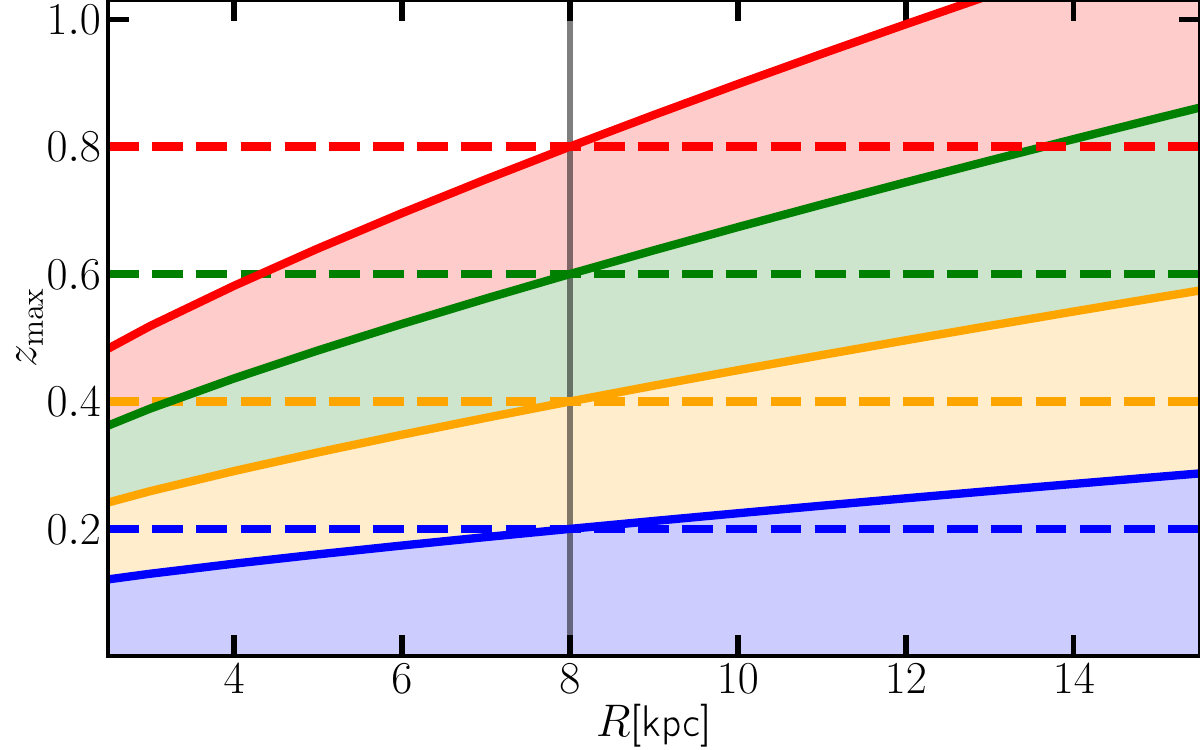}
    \caption{The selections of stars via $z_\mathrm{max}$ as a function of $R$. The dashed lines correspond to the upper limit of a constant selection (which was applied in \Fref{fig:ecc_R_kord_unscaled}). The coloured background (and the solid lines as upper limits) correspond to the selection gauged with the Milky Way potential \citep{McMillan2017}.
    This $z_\mathrm{max}$ was applied in \Fref{fig:ecc_R_kord}.}
    \label{fig:norm_zmax}
\end{figure}

\begin{figure}
    \centering
    \includegraphics[width=0.47\textwidth]{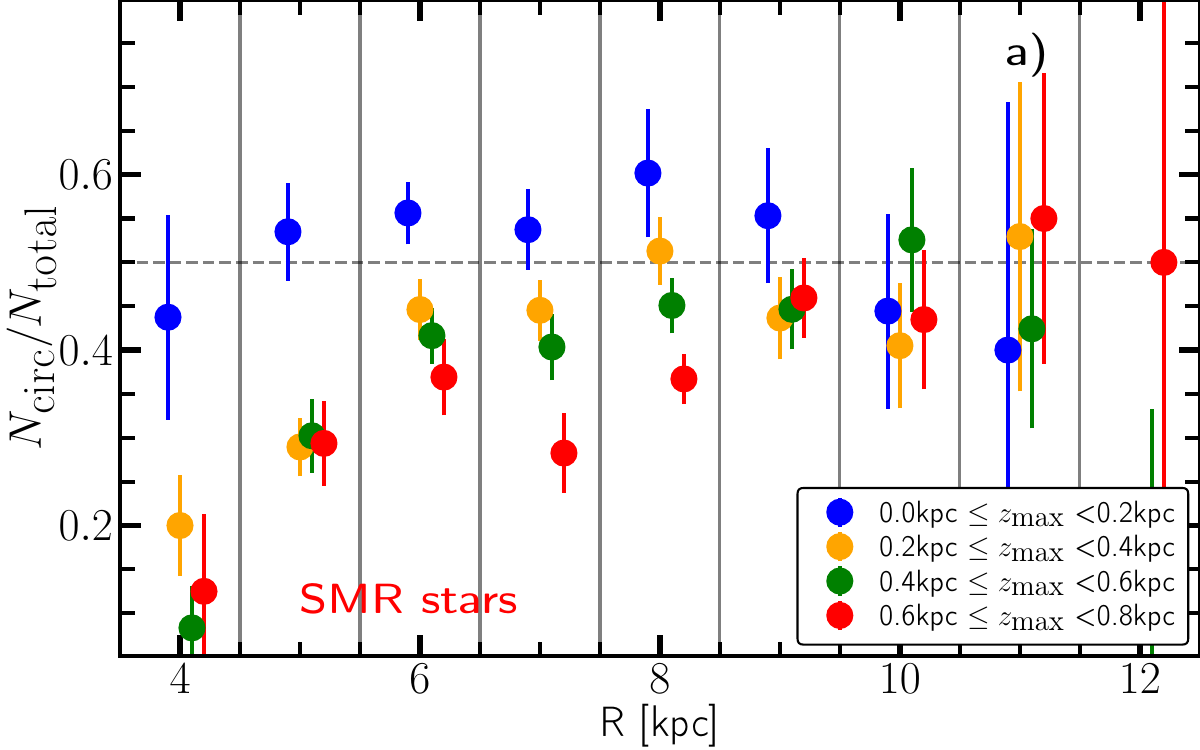}
    \includegraphics[width=0.47\textwidth]{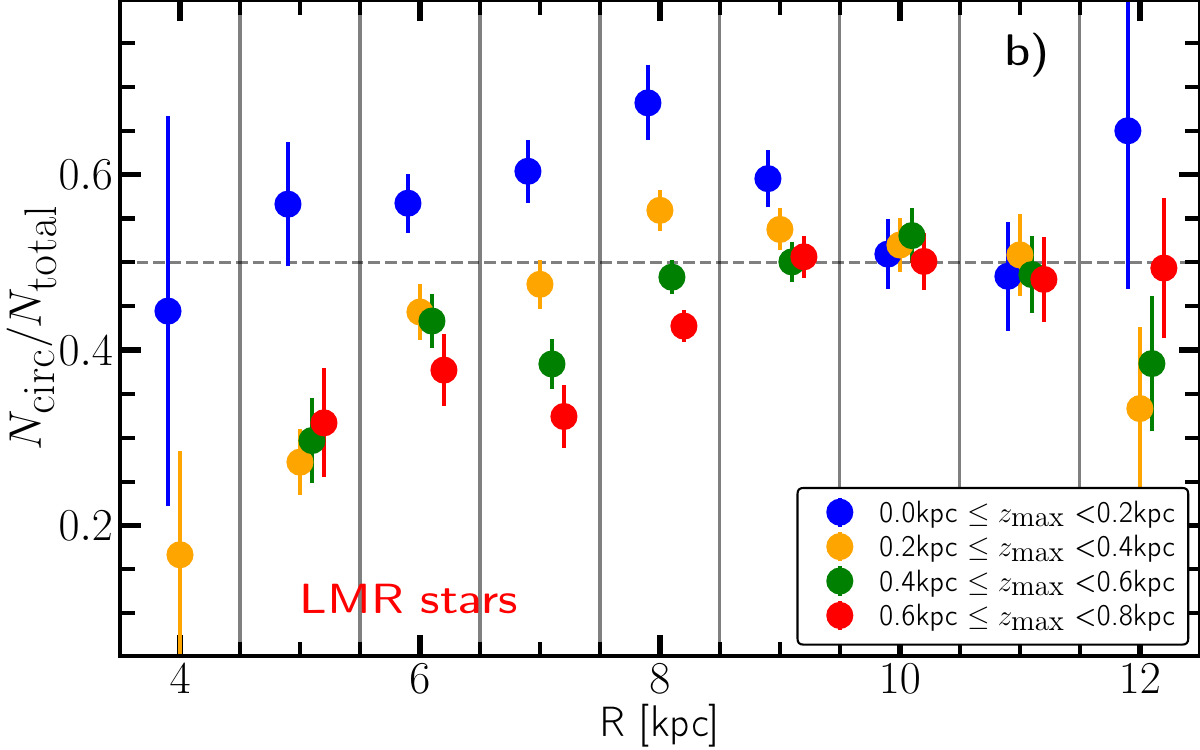}
    \caption{Fraction of stars on circular orbits split into $R$-bins. The a) SMR and b) LMR selections differ from each other especially in the outer Galaxy. In these diagrams, the $z_\mathrm{max}$ selection is constant, i.e.\ sliced into regions of $\SI{200}{pc}$ around the plane.}
    \label{fig:ecc_R_kord_unscaled}
\end{figure}

Therefore, we introduce a new quantity, the effective $z$ amplitude $\zeta$, which is calculated in the following way:
\begin{align}
    \zeta = z_\mathrm{max}\times g(R) \textrm{, with } g=\frac{P_\mathrm{M17}(R, z=0)}{P_\mathrm{M17}(R=\SI{8}{kpc}, z=0)},
\end{align}
where we use the Milky Way potential $P(R, z)$ from \citet{McMillan2017} through \textsc{galpy} \citep{galpy15}. Note that this is constructed to be equivalent to $z_\mathrm{max}$ at $\sim \SI{8}{kpc}$.
In \Fref{fig:norm_zmax} we show the lines of constant $z_\mathrm{max}$ (dashed lines) and $\zeta$ (solid lines) used over the full range of Galactocentric radii. The constant $z_\mathrm{max}$ selection is applied in \Fref{fig:ecc_R_kord_unscaled}.

In \Fref{fig:ecc_R_kord_unscaled} we show the equivalent of \Fref{fig:ecc_R_kord} using a selection via $z_\mathrm{max}$ (no modifications). We can see that qualitatively the same trends apply in both figures. However, this sort of modification is still important as it otherwise might lead us to conclude that some of our results are caused by variations in the potential rather than actual trends in the red giant populations.

\begin{table*}
\caption{Number of SMR stars in each bin of panel a) in \Fref{fig:ecc_R_kord}. The top row displays the $R$ bins in $\SI{}{kpc}$ and the leftmost column the $\zeta$ in $\SI{}{kpc}$.}
\begin{center}
\vspace{0cm}
\begin{tabular}{c|ccccccccc}
    $\zeta$ & $[3.5, 4.5]$ & $[4.5, 5.5]$ & $[5.5, 6.5]$ & $[6.5, 7.5]$ & $[7.5, 8.5]$ & $[8.5, 9.5]$ & $[9.5, 10.5]$ & $[10.5, 11.5]$ & $[11.5, 12.5]$ \\
    \hline
    $[0, 0.2]$ & 0 & 11 & 118 & 372 & 233 & 113 & 106 & 41 & 9 \\
    $[0.2, 0.4]$ & 2 & 48 & 212 & 356 & 341 & 349 & 210 & 95 & 23 \\
    $[0.4, 0.6]$ & 5 & 45 & 177 & 316 & 285 & 448 & 236 & 77 & 36 \\
    $[0.6, 0.8]$ & 9 & 22 & 128 & 294 & 159 & 444 & 217 & 84 & 21 \\
    \hline
\end{tabular}
\end{center}
    \label{tab:number_ecc_zeta_a}
\end{table*}

\begin{table*}
\caption{Number of LMR stars in each bin of panel b) in \Fref{fig:ecc_R_kord}. The top row displays the $R$ bins in $\SI{}{kpc}$ and the leftmost column the $\zeta$ in $\SI{}{kpc}$.}
\begin{center}
\vspace{0cm}
\begin{tabular}{c|ccccccccc}
    $\zeta$ & $[3.5, 4.5]$ & $[4.5, 5.5]$ & $[5.5, 6.5]$ & $[6.5, 7.5]$ & $[7.5, 8.5]$ & $[8.5, 9.5]$ & $[9.5, 10.5]$ & $[10.5, 11.5]$ & $[11.5, 12.5]$ \\
    \hline
    $[0, 0.2]$ & 4 & 73 & 433 & 438 & 374 & 626 & 371 & 164 & 27 \\
    $[0.2, 0.4]$ & 10 & 163 & 403 & 558 & 1003 & 997 & 639 & 303 & 58 \\
    $[0.4, 0.6]$ & 9 & 125 & 377 & 500 & 1333 & 1027 & 559 & 251 & 95 \\
    $[0.6, 0.8]$ & 5 & 90 & 335 & 279 & 1301 & 883 & 491 & 219 & 79 \\
    \hline
\end{tabular}
\end{center}
    \label{tab:number_ecc_zeta_b}
\end{table*}

\bsp	
\label{lastpage}
\end{CJK*}
\end{document}